\documentclass[11pt,a4paper]{article}
\usepackage{fullpage}
\usepackage{authblk}

\usepackage{mathrsfs}
\usepackage{amsfonts}
\usepackage{amssymb}
\usepackage{bm}
\usepackage{multirow,bigstrut}
\usepackage{threeparttable}
\usepackage{enumerate}
\usepackage{tikz}
\usepackage{graphicx}
\usepackage{stfloats}
\usepackage{array}
\usepackage{amsmath}
\usepackage[colorlinks,
            linkcolor=red,
            anchorcolor=blue,
            citecolor=green
            ]{hyperref}

\newtheorem{theorem}{Theorem}[section]

\newtheorem{remark}[theorem]{Remark}

\newtheorem{lemma}[theorem]{Lemma}
\newtheorem{example}[theorem]{Example}

\numberwithin{equation}{section}

\def \H {\mathbb{H}}
\def \T {\mathbb{T}}
\def \C {\mathbb{C}}

\def\i { \bm {i}}

\def\sgn{\mbox{sgn}}

\def \T {\mathbb{T}}
\def \Z {\mathbb{Z}}
\def\sgn{\mbox{sgn}}
\def \A {\mathbf{A}}
\def \C {\mathbf{C}}
\def \D {\mathbf{D}}
\def \E {\mathbf{E}}
\def \F {\mathbf{F}}
\def \U {\mathbf{U}}
\def \V {\mathbf{V}}

\newcommand{\norm}[1]{\left\lVert#1\right\rVert}
\newcommand{\abs}[1]{\left|#1\right|}

\newenvironment{nproof}{\noindent{\em \textbf{Proof.}}}{\quad \hfill$\Box$\vspace{2ex}}

\title{FFT Multichannel Interpolation and Application to Image Super-resolution}
	\author{Dong Cheng\thanks{chengdong720@163.com} }
	
	\author{Kit Ian Kou\thanks{kikou@umac.mo}}
	
	\affil{\normalsize{Department of Mathematics, Faculty of Science and Technology, University of Macau, Macao, China}}
	
	\date{}

\begin{document}
  \maketitle
\begin{abstract}
\normalsize
This paper presents an innovative set of tools to support a methodology for the multichannel interpolation (MCI) of a discrete signal. It is shown that a bandlimited signal $f$ can be exactly reconstructed from finite samples of $g_k$ ($1\leq k\leq M$) which are the responses of $M$ linear systems with input $f$.  The proposed interpolation can also be applied to approximate non-bandlimited signals. Quantitative error is analyzed to ensure its effectiveness in approximating non-bandlimited signals and its Hilbert transform. Based on the FFT technique, a fast algorithm which brings high computational efficiency and reliability for MCI is  presented. The standout performance of MCI is illustrated by several simulations. Additionally, the proposed interpolation is applied to the single image super-resolution (SISR). Its superior performance in accuracy and speed of SISR is demonstrated by the experimental studies. Our results are compared qualitatively and quantitatively with the state-of-the-art methods in image upsampling and reconstruction by using the standard measurement criteria.
\end{abstract}


\begin{msc}
 42A15,  94A12, 65T50.
\end{msc}

\section{Introduction}\label{S1}

{I}nterpolation by simple functions, such as trigonometric functions or rational functions, is an important mathematical technique used in physics and engineering sciences. It deals with the problem of reconstructing or approximating the continuous signals from a series of discrete points.
An ideal interpolation usually tells us that a continuous signal can be exactly and uniquely reconstructed from the discrete points if it satisfies some suitable conditions, for instance, bandlimited in Fourier domain. Such an interpolation is also referred to as a sampling theorem in signal processing \cite{zayed1993advances,torres2006sampling}.
In general, sampling formulas interpolate the given data even if the specific conditions for perfect reconstruction cannot be met. Therefore the error analysis of such sampling formulas is of great importance, because the recovered signal does not satisfy the conditions for ideal interpolation in most circumstances.
Due to the wide range applications of interpolation, finding new interpolation or sampling formulas with error  estimations as well as  developing their fast algorithms for implementation  have received considerable attentions in recently years, see for instance \cite{sharma2007papoulis,selva2009functionally,liu2010new,zhang2015sampling,cheng2017novel}.

Most of the classical sampling formulas have centered around reconstructing a signal from its own samples \cite{zayed1993advances,selva2015fft}. In fact, reconstructing a signal from  data other than the samples of the original signal is possible. Papoulis \cite{papoulis1977generalized} first proposed generalized sampling expansion (GSE) for bandlimited signals defined on real line $\mathbb{R}$. The GSE indicates that a  $\sigma$-bandlimited signal $f$ can be reconstructed from the samples of output signals of $M$ linear systems. That is, one can reconstruct $f$  from the samples $g_1(nT),\dots,g_M(nT)$ of the output signals
\begin{equation*}
g_m(t)=\frac{1}{\sqrt{2\pi}}\int_{-\sigma}^{\sigma}F(\omega)H_m(\omega)e^{\i t \omega}d\omega,~~m=1,\dots,M
\end{equation*}
where $F$ is Fourier transform (FT) of $f$ and $H_1,\dots,H_M$ are the system functions and $T=M\pi/\sigma$.
The study of GSE  has been extended in various directions.  Cheung \cite{cheung1993multidimensional}
introduced   GSE  for real-valued  multidimensional signals associated with Fourier transform, while Wei, Ran and Li \cite{wei2010generalized,wei2011multichannel}
presented the GSE with generalized integral transformation, such as fractional Fourier transform (FrFT) and linear canonical transform (LCT). Some new sampling models in FrFT domain, such as shift-invariant spaces model \cite{zhao2018generalized} and multiple sampling rates model \cite{wei2016generalized} were discussed.  In \cite{chengkou2018generalized}, the authors studied GSE for quaternion-valued signals associated with quaternion Fourier transform (QFT). Importantly, the applications based on GSE as well as its generalizations  have been widely conducted \cite{sharma2007papoulis,wei2016generalized,li2014image}.

In a real application, the actual signal's length is generally limited. The uncertainty principle states that a time-limited signal can not be bandlimited simultaneously in FT   domain. However, it is known that the classical Shannon sampling theorem and GSE are aimed at reconstructing bandlimited signals in  FT or FrFT domain. Therefore, a finite duration signal is commonly assumed to be a part of periodic signal in practice. Accordingly, there are certain studies concerning the interpolation theory of finite duration signals. A sampling theorem for  trigonometric polynomials was first introduced by  Goldman \cite{goldman1953information}.  The author in \cite{ogawa1989generalized} proposed   pseudo-orthogonal bases for reconstructing   signals defined on  finite interval. In a series of papers \cite{schanze1995sinc,candocia1998comments,dooley2000notes}, researchers extensively discussed the sinc interpolation of discrete cyclic  signals and they also derived several equivalent interpolation formulas in distinct forms.
As  an extension of  cyclic sinc interpolation, decomposing a finite duration signal in a basis of shifted and scaled versions of a generating function was studied in \cite{jacob2002sampling}. Moreover, they further presented  an error analysis for their  approximation method.    Recently, the non-uniform sampling theorems for finite duration signals were also presented \cite{Margolis2008nonuniform,xiao2013sampling}.

In this paper, a multichannel interpolation for finite duration signals is studied. We derive a general interpolation formula that depends on the transfer functions of the linear systems. The formula  bears a resemblance to the classical GSE defined on real line. Nevertheless,   the recipe of derivation is different from the traditional one, and moreover,  the proposed formula  is  given   by  a finite summation,  as opposed to a infinite summation in the traditional case. Not only the theoretical error analysis but also the numerical simulations are provided to   show the effectiveness of MCI in signal reconstruction. Since MCI is a novel interpolation method with  high accuracy and   efficiency, we also apply it to image super-resolution. 

Single image  super-resolution (SISR) is of  importance in many personal, medical and industrial imaging applications \cite{li2014image,lehmann1999survey,thevenaz2000interpolation,battiato2002locally}. 
The existing SISR techniques roughly fall into three categories: the interpolation-based methods \cite{zhang2006edge,wei2016generalized}, the learning-based methods \cite{takeda2007kernel,timofte2013anchored,dong2016image,kim2016accurate,wang2017fast,wang2017single,tai2017image} and the reconstruction-based methods \cite{yang2010image,dong2011image,dong2013nonlocally,zeyde2010single,chang2018single1}.  
The learning-based and reconstruction-based SISR techniques have been recognized as   effective approaches to produce   high-quality images with fine details from  low-resolution inputs. It is known that, however, the performance of the learning-based methods depends much on  the similarity between the images for training and for testing,  while the performance of the reconstruction-based SISR methods relies on the reasonability  of the incorporated prior information. Inspired by the success of the learning-based SISR methods,   the authors in \cite{chang2018single} proposed  a novel reconstruction-based  SISR method (called CRNS algorithm) which     utilizes the complementary advantages
of both the learning-based and the reconstruction-based methods. That is, CRNS algorithm takes both the external and internal priors into consideration to improve SISR results.

In spite of the remarkable performance of  learning-based and reconstruction-based methods in SISR, the interpolation-based methods are also widely used to produce high resolution (HR) images for their computational efficiency. From Table \ref{averagetime}, our MCI method holds over $300$ times faster than the CRNS method. Besides producing HR images, the interpolation-based methods  are commonly incorporated into other methods to improve HR results. In \cite{wang2014fast}, the authors proposed a two-scale approach to recover HR images by interpolation and reconstruction respectively. The gradient profile model in \cite{sun2011gradient} is trained from LR and   HR image pairs, where the initial HR images are  produced by interpolation-based methods. Therefore, it is convinced that a good interpolation-based   method can be useful to SISR. 
Recently, the application of   GSE associated with FrFT        to SISR  was investigated \cite{wei2016generalized}.  Compared with the classical image interpolation algorithms  such as Lanczos and bicubic, the GSE-based algorithm tends to
produce images with less blur and good contrast. However,
the conventional Shannon sampling theorem \cite{zayed1993advances} and GSE  associated with FT or FrFT \cite{papoulis1977generalized,wei2016generalized} are involved in infinite number of sample values  spreading over  whole real line and the interpolation functions have  infinite  duration. Once they are applied to  SISR, the truncation errors are  inevitable. Fortunately,  there is no problem with truncation error for MCI. Besides, the MCI-based SISR algorithm can preserve  lots of information of original image for reshaped image, in view of the proposed MCI   makes good use of multifaceted information such as  first derivative (which may include edge information of image) and second derivative (which may include detail information of image). It will be shown that the MCI-based algorithm can produce better SISR results  than GSE-based algorithm.
Moreover,   the FFT-based  implementation  of MCI   makes    the proposed   algorithm  very  fast (see Table \ref{averagetime}). To validate the performance of our SISR method, we further  combine MCI with displacement field (FD) method such that the produced HR images can preserve sharp edges. The experiments show that the FD-based method achieves a significant improvement by introducing MCI.

In summary,  the contributions of this paper are highlighted as follows:
\begin{enumerate}
	\item  We propose a multichannel interpolation  for finite duration signals (MCI). The proposed MCI is  capable of generating various useful interpolation formulas by selecting suitable parameters according to   the types and the amount of collected data.
	In addition, it would restore the original signal $f$ and some integral transformations (such as Hilbert transform) of $f$.
	\item Based on FFT, a fast algorithm  which brings high computational efficiency and reliability  for  MCI is also presented.
	\item  Two questions naturally arise when using a sampling or interpolation  formula to reconstruct a non-bandlimited signal. One is whether  the set of original samples stays unchanged after reconstruction, namely, whether the interpolation consistency holds. The other one is whether the error of  imperfect reconstruction can be estimated.  To the authors' knowledge, these two issues have not been addressed for the classical GSE in the  literature. By contrast,  error analysis arising in  reconstructing non-bandlimited signals by the proposed MCI is studied. Moreover, the corresponding interpolation consistency is also proved.
	\item  The proposed MCI is applied to single image super-resolution reconstruction (SISR). Its main advantage is making good use of multifaceted information of  image, so that the reconstructed image retains lots of  information of the original image. Moreover, the interpolation consistency and the untruncated  implementation of MCI can reduce the reconstruction errors.  The superior performance of  the proposed algorithm in accuracy and  speed of   SISR is shown by several experimental simulations.
\end{enumerate}

The rest of the paper is organized as follows: Section \ref{S2} recalls some preliminaries of  Fourier series. Section \ref{S3} formulates MCI and presents some examples to illustrate how to use MCI flexibly. The error analysis and interpolation consistency are drawn in Section \ref{S4}.  In Section \ref{S5}, the effectiveness of the proposed MCI  for approximating signals is demonstrated by several numerical  examples  and the application of MCI to single image super-resolution reconstruction is also addressed. Finally, conclusions are made in Section \ref{S6}.


\section{Preliminaries}\label{S2}
This part   recalls some preparatory knowledge  of Fourier series (see e.g. \cite{folland1992fourier}). Throughout the paper,   the set of real numbers, integers and positive integers  are denoted by $\mathbb{R}$, $\mathbb{Z}$ and $\mathbb{Z}^+$ respectively.
Without loss of generality, we restrict attention to the  signals defined on unit circle $\T:=[0,2\pi)$.

Let $L^p(\T)$   be the totality of functions $f(t)$ such that
\begin{equation*}
\norm{f}_p:= \left(\frac{1}{2\pi} \int_{\T}|f(t)|^pdt\right)^{\frac{1}{p}}<\infty,
\end{equation*}
and $l^{p}$ be the  sequence space defined by
\begin{equation*}
l_p:=  \left\{\{x(n)\}: \sum_{n\in \mathbb{Z}}\abs{x(n)}^p<\infty\right\}
\end{equation*}
For $f\in L^2(\T)\subset L^1(\T)$, it can be written as
\begin{equation}\label{function}
f(t)=\sum_{n\in \mathbb{Z}} a(n) e^{\i nt}
\end{equation}
with $\sum_{n\in \mathbb{Z}} \abs{a(n)}^2<\infty$, where the Fourier series is convergent to $f$ in $L^2$ norm. It is well known that $L^2(\T)$ is a  Hilbert space  with the inner product
\begin{equation*}
(f,h):= \frac{1}{2\pi} \int_{\T}f(t)\overline{h(t)}dt,~ ~~\forall f, h \in L^2(\T).
\end{equation*}
If $f,h\in L^2(\T)$  and their Fourier coefficients are respectively given by
\begin{equation*}
f\sim  a(n) ,~~~~h\sim b(n) .
\end{equation*}
By H{\"o}lder inequality, we have $\{a(n) b(n)\},\{a(n) \overline{b(n)} \}\in l^1$ and
the general version of Parseval's identity is of the form:
\begin{equation*}
(f,h) =\sum_{n\in \Z}a(n) \overline{b(n)}.
\end{equation*}
Moreover, the convolution theorem gives
\begin{equation}\label{conv}
(f*h)(t)  :=\frac{1}{2\pi} \int_{\T}f(s)h(t-s)ds=\sum_{n\in \Z}a(n) b(n) e^{\i nt}.
\end{equation}

The analytic signal  defined by the linear combination of original signal and its (circular) Hilbert transform  is regarded as an useful representation from which the phase, energy and frequency may be estimated \cite{gabor1946theory,cohen1995time}.
The circular Hilbert transform \cite{king2009hilbert,mo2015afd} for   signal $f$ is given by
\begin{equation*}
\begin{split}
\mathcal{H}f(t)  & :=   \sum_{n\in \mathbb{Z}}(-\i\sgn(n))a(n) e^{\i n t} \\
&  =\frac{1}{2\pi}\mathrm{p.v.}\int_{\T}
f(s)\cot\left(\frac{t-s}{2}\right)ds,
\end{split}
\end{equation*}
where $\mathrm{p.v.}$ is the abbreviation of Cauchy principal value and  $\sgn$ is the signum function taking values $1$, $-1$ or $0$ for $n>0$, $n<0$ or $n=0$ respectively. By definition, we have
\begin{equation}\label{Hsquare}
\mathcal{H}^2f(t)= -f(t)+a(0).
\end{equation}

We will first concentrate on reconstruction problem of finite order  trigonometric polynomials. To maintain consistent terminology with the classical case, in what follows, a finite order  trigonometric polynomial is called a bandlimited   signal. Specifically, a  signal  $f(t)$  is said to be bandlimited  if   its sequence of Fourier  coefficients possesses  finite nonzero elements. Let $\mathbf{N}=(N_1,N_2)\in \mathbb{Z}^2$, in the sequel we denote by  $B_{\mathbf{N}}$ the totality of bandlimited signals with the following form:
\begin{equation}\label{bandsignal}
f(t)=\sum_{n\in I^{\mathbf{N}}}a(n)e^{\i nt}  ,~~~I^{\mathbf{N}}=\{n: N_1\leq n \leq N_2\}.
\end{equation}

\section{Formulation of MCI}\label{S3}

Let $N_1,N_2\in \mathbb{Z}$, $M\in \mathbb{Z}^+$ and assume that $N_2-N_1+1$ is   divisible by $M$, namely, $\frac{N_2-N_1+1}{M}=L\in \mathbb{Z}^+$.
We cut the set of integers into pieces    for convenience.  Let us set
\begin{equation*}
I_k=\{n: N_1+(k-1)L\leq n\leq N_1+kL-1\},~~J_k= \bigcup_{l=k+1}^{M+k}I_l.
\end{equation*}
Then we have $ I^{\mathbf{N}}=\bigcup_{k=1}^M I_k=J_0$ and $ \mathbb{Z}=\bigcup_k I_k $.

For $1\leq m \leq M$, let
\begin{align}
h_m(t)  &=\sum_{n\in \Z}b_{m}(n)e^{\i  nt}, \label{sysfunction}  \\
g_m(t)   & = (f * h_m)(t) =  \frac{1}{2\pi} \int_{\T}f(s)h_m(t-s)ds.\nonumber
\end{align}
It follows from (\ref{conv})  that $$g_m(t)=\sum_{n\in \Z} c_{m}(n)e^{\i  nt},$$
where $c_m(n)= a(n)b_m(n)$. We particularly mention that the series (\ref{sysfunction}) may not be convergent in general.  Nevertheless, $g_m(t)$ is well defined when $\{c_m(n)\}\in l^1$.  In this case, $h_m(t)$ may be regarded as a distribution.

The proposed MCI is to reconstruct $f(t)$ from the samples of $g_1(t),g_2(t),\dots,g_M(t)$. To achieve this, it is natural to expect that the simultaneous sampling of $M$  signals   will reduces  sampling rate by $1/M$. we shall note that the signal expressed as Eq. (\ref{bandsignal}) may be represented by a shorter length of summation as long as we introduce the following vectors.

For $n\in I_1$, let
\begin{align*}
& \A_n   = \left[a(n),a ({n+L}),a({n+2 L}),\cdots,a({n+(M-1)L})\right], \\
& \E_n(t)   = \left[e^{\i nt} ,e^{\i(n+L)t},e^{\i(n+2 L)t},\cdots,e^{\i(n+(M-1)L)t}\right]^{\rm T}.
\end{align*}
It follows that
\begin{equation}\label{fmatrix}
\begin{split}
f(t) & =\sum_{n\in I^{\mathbf{N}}}a(n)e^{\i nt} =\sum_{n=I_1}\sum_{k=0}^{M-1}a({n+k L}) e^{\i (n+k L)t} \\
&  = \sum_{n=I_1}\A_n \E_n(t).
\end{split}
\end{equation}
Similar considerations applying to $g_m(t)$, we have
\begin{equation*}
g_m(t)= \sum_{n\in I_1}\C_{m,n}  \E_n(t),
\end{equation*}
where
$$\C_{m,n}  = \left[c(n),c ({n+L}),c({n+2 L}),\cdots,c({n+(M-1)L})\right].$$
Owing to the periodicity of $\E_n(t)$, we easily obtain
\begin{equation*}
\E_n(\frac{2\pi p}{L})=e^{\i n\frac{2\pi p}{L}}[1,1,\cdots,1]
\end{equation*}
for every $0 \leq p \leq L-1$. This leads to a simple expression for samples of $g_m(t)$, that is
\begin{equation}\label{sumformgm}
\begin{split}
g_m(\frac{2\pi p}{L}) &  =\sum_{n\in I_1} \C_{m,n}\E_n(\frac{2\pi p}{L}) \\
& = \sum_{n\in I_1}e^{\i n\frac{2\pi p}{L}} \sum_{k=0}^{M-1}c_{m}(n+k L)\\
& = \sum_{n\in I_1} d_{m}(n) e^{\i n\frac{2\pi p}{L}},
\end{split}
\end{equation}
where $ d_{m}(n)=\sum_{k=0}^{M-1}c_{m}(n+kL)$. This indicates that  $g_m(\frac{2\pi p}{L})$ can be computed by taking discrete Fourier transform of $d_m(n)$ with respect to $n$. It follows that we could be able to express the samples of $g_m(t)$   in terms of the  DFT matrix  or its inverse matrix.
\begin{lemma}\label{matrixsamplegm}
	Let  $t_p=\frac{2\pi p}{L}$ and
	$\widetilde{\D}_m= \left[d_{m}(N_1),d_{m}(N_1+1),\cdots,d_{m}(N_1+L-1)\right].$
	There is a matrix representation for samples of $g_m(t)$ in terms of the inverse DFT matrix. That is,
	\begin{equation}\label{matrixgm}
	\frac{1}{L} \left[g_m(t_0),g_m(t_1),\cdots,g_m(t_{L-1})\right]=\widetilde{\D}_m \F_L^{-1} \U_L^{-1},
	\end{equation}
	where $\F_L$ is  the $L$-th order DFT matrix
	\begin{equation}\label{DFTmatrix}
	\F_L=\begin{bmatrix}
	\omega^0 & \omega^0& \omega ^0&\cdots &\omega^0\\
	\omega^0 & \omega^1& \omega^2 &\cdots &\omega^{L-1}\\
	\omega^0 & \omega^2& \omega^4 &\cdots &\omega^{2(L-1)}\\
	\vdots&\vdots&\vdots& \ddots & \vdots \\
	\omega^0 & \omega^{L-1}& \omega^{2(L-1)} &\cdots &\omega^{(L-1)^2}\\
	\end{bmatrix}
	\end{equation}
	with $\omega=e^{ {-2\pi\i}/{L}}$ and $\U_L$ is a diagonal matrix
	\begin{equation}\label{shiftmatrix}
	\U_L= \begin{bmatrix}
	\omega^0\\
	&\omega^{N_1}& &\text{{\huge 0}}\\
	& & \omega^{2{N_1}} \\
	& \text{{\huge0}} & & \ddots\\
	& & & & \omega^{(L-1){N_1}}
	\end{bmatrix}.
	\end{equation}
\end{lemma}
\begin{nproof}
	From   Eq. (\ref{sumformgm}), we can  rewrite $g_m(t_p)$ as
	\begin{equation*}
	\begin{bmatrix}
	d_{m}(N_1)\\
	d_{m}(N_1+1) \\
	\vdots\\
	d_{m}(N_1+L-1)
	\end{bmatrix}^{\rm T}
	\begin{bmatrix}
	e^{\i N_1 \frac{2\pi p}{L}} \\
	e^{\i(1+ N_1) \frac{2\pi p}{L}} \\
	\vdots\\
	e^{\i(L-1+ N_1) \frac{2\pi p}{L}}
	\end{bmatrix}.
	\end{equation*}
	Note that $\omega=e^{ {-2\pi\i}/{L}}$, $\U_L ^{-1}=\overline{\U_L} $ and $\F_{L}^{-1}=\frac{1}{L} \overline{\F_L}$, it follows that
	\begin{equation*}
	\begin{split}
	&\frac{1}{L} \left[ e^{\i N_1 \frac{2\pi p}{L}}, e^{\i(1+ N_1) \frac{2\pi p}{L}},\cdots,  e^{\i(L-1+ N_1) \frac{2\pi p}{L}}\right]^{\rm T} \\
	= & \text{Product of} ~\F_L^{-1}~ \text{and} ~(p+1)\text{-th column of}~ \U_L^{-1}.
	\end{split}
	\end{equation*}
	Hence we immediately obtain Eq. (\ref{matrixgm}) which completes the proof.
\end{nproof}

Following the definition of $\A_n$, for $n\in I_1$,  we further set $M$ by $M$ matrix
\begin{equation}\label{defH}
\H_n= \left[ b_k(n+jL-L)\right]_{jk},
\end{equation}
which means  the $jk$-th element of $\H_n$ is $ b_k(n+jL-L)$.
Suppose that $\H_n$ is invertible for every $n \in  I_1$ and we denote the inverse matrix as
\begin{equation*}
\H_n^{-1}= \begin{bmatrix}
q_{11} (n) & q_{12} (n)&\cdots &q_{1M}(n)\\
q_{21}(n) & q_{22}(n) &\cdots &q_{2M} (n)\\
\vdots&\vdots& ~ & \vdots \\
q_{M1} (n)& q_{M2}(n) &\cdots &q_{MM}(n)
\end{bmatrix}.
\end{equation*}
Next we  use the elements of $\H_n^{-1}$ to construct the interpolation functions. Let
\begin{equation}\label{defrm}
r_{m}(n)= \begin{cases}
q_{mk}(n+L-kL),&  \text{if}~n\in I_k, ~k=1,2,\cdots,M ,  \\
0 & \text{if}~ n\notin I^{\mathbf{N}},
\end{cases}
\end{equation}
and define
\begin{equation}\label{ym}
y_m(t)= \sum_{n\in I^{\mathbf{N}}}r_{m}(n)e^{\i n t}.
\end{equation}
As with most conventional bandlimited interpolation methods, the interpolation functions of  MCI are generated from some fixed  functions by translations. The following result provides  a matrix representation of shifted functions generated by $y_m(t)$.
\begin{lemma}\label{ymt-tp}
	Let $t_p=\frac{2\pi p}{L}$ and $y_m(t)$ be defined by (\ref{ym}). The shifted functions of $y_m(t)$ can be expressed by
	\begin{equation}\label{ymmatrix}
	\left[y_m(t-t_0),y_m(t-t_1),\cdots,y_m(t-t_{L-1})\right]=\widetilde{\V}_m(t)  {\F_L} \U_L
	\end{equation}
	where $\F_L, \U_L$ are given respectively  by (\ref{DFTmatrix}), (\ref{shiftmatrix}) and $\widetilde{\V}_m(t)$ is defined as
	\begin{equation*}
	\widetilde{\V}_m(t)=\left[v_{m,N_1}(t), v_{m,N_1+1}(t),\cdots, v_{m,L+N_1-1}(t)\right]
	\end{equation*}
	with $v_{m,n}(t)=\sum_{k=1}^M  q_{mk}(n) e^{\i (n+kL-L)t}$, for  $n \in I_1$.
\end{lemma}
\begin{nproof}
	By the definition of $y_m$  and rearranging the summation terms of Eq. (\ref{ym}), we get
	\begin{equation*}
	\begin{split}
	& y_m(t-t_p)   \\
	=  &  \sum_{n \in I^{\mathbf{N}}}r_{m}(n)e^{\i n (t-\frac{2\pi p }{L})} \\
	= &  \sum_{k=1}^M  \sum_{n\in I_1} r_{m}(n+kL-L) e^{\i (n+kL-L)(t-\frac{2\pi p }{L})}\\
	=&  \sum_{n\in I_1}\left( \sum_{k=1}^M r_{m}(n+kL-L) e^{\i (n+kL-L)t}\right) e^{-\i n\frac{2\pi p}{L}}.
	\end{split}
	\end{equation*}
	From the piecewise defined $r_m(n)$ in Eq. (\ref{defrm}), it is easy to verify that
	$ r_{m}(n+kL-L) =q_{mk}(n)$
	for all $ n \in I_1$ and $1\leq k \leq M$. It follows that $y_m(t-t_p)$ can be rewritten as
	\begin{equation}\label{transformym}
	y_m(t-t_p) = \sum_{n\in I_1}  v_{m,n}(t) e^{-\i n\frac{2\pi p}{L}}.
	\end{equation}
	In view of  the resemblance of Eq. (\ref{transformym}) and Eq. (\ref{sumformgm}) and by similar arguments to Lemma \ref{matrixsamplegm}, we  immediately have Eq. (\ref{ymmatrix}) which completes the proof.
\end{nproof}

Until now, in addition to representing the sampled values of $g_m(t)$  by $\widetilde{\D}_m$,  the interpolation functions $ y_m(t-t_p)$ ($1\leq m \leq M$, $0\leq p \leq L-1$) have been    constructed from the elements of $\H_n^{-1}$.
we shall now make  a connection between $f(t)$ and $ y_m(t-t_p)$. To achieve  this, let
\begin{align*}
\D_n&=\left[d_{1}(n),d_2(n), d_{3}(n),\cdots,d_{M}(n)\right],\\
\V_n(t)&  = \left[v_{1,n}(t) ,v_{2,n}(t),v_{3,n}(t),\cdots,v_{M,n}(t)\right].
\end{align*}
By straightforward computations, we get
\begin{align}
\A_n \H_n &=\D_n, \label{matrixeqn1} \\
\H_n\left[\V_n(t)\right]^{\rm T} &  = \E_n (t).  \label{matrixeqn2}
\end{align}
Having introduced the above notations,  we are in position to show our main result which is referred to as multichannel interpolation   (MCI).
\begin{theorem}\label{MCIeq}
	Let $f\in B_{\mathbf{N}}$ and   $g_m$, $\H_n$ be given above. If  $\H_n$ is  invertible for all
	$n \in  I_1$, then
	\begin{equation}\label{drictexpression}
	f(t)=  \frac{1}{L} \sum_{m=1}^{M} \sum_{p=0}^{L-1}g_m(t_p) y_m(t-t_p)
	\end{equation}
	where $t_p=\frac{2\pi p}{L}$ and $y_m$ is given by  Eq. (\ref{ym}).
\end{theorem}
\begin{nproof}
	Multiplying  Eq. (\ref{matrixeqn2})  by $\A_n$  and using Eq.   (\ref{matrixeqn1}),  the  expression (\ref{bandsignal}) can be rewritten as
	\begin{align}
	f(t)&  =\sum_{n\in I_1}\D_n \left[\V_n(t)\right]^{\rm T} =\sum_{n\in I_1}\sum_{m=1}^{M} d_{m}(n) v_{m,n}(t)\nonumber \\
	&= \sum_{m=1}^{M}   \sum_{n\in I_1}d_{m}(n) v_{m,n}(t)=\sum_{m=1}^{M} \widetilde{\D}_m [\widetilde{\V}_m(t)]^{\rm T}    \nonumber  \\
	&= \sum_{m=1}^{M} \widetilde{\D}_m \F_L^{-1}   \U_L^{-1} \U_L {\F_L}[\widetilde{\V}_m(t)]^{\rm T} \label{DFTexpression}\\
	&=\frac{1}{L} \sum_{m=1}^{M} \sum_{p=0}^{L-1}g_m(t_p) y_m(t-t_p)\nonumber
	\end{align}
	The last equality is a consequence of Lemma \ref{matrixsamplegm} and \ref{ymt-tp}. The proof is complete.
\end{nproof}
\begin{remark}\label{aequal0}
	In general,   $\H_n$ has to be invertible for every $n\in I_1$  because the definition of $v_{m,n}(t)$  depends  on $q_{mk}(n)$ ($1\leq k \leq M$) which are the elements of  $\H_n^{-1}$. However, if there is  an index set $\Lambda\subset I_1$ such that $\A_n=\mathbf{0}$ for every $n\in \Lambda$, then $\H_n$ is not necessary to be invertible for $n\in \Lambda$. This is due to the fact that the zero terms of  (\ref{fmatrix}) may be removed from the summation. In this case, we just need to set $q_{mk}(n)=0$ ($1\leq m, k \leq M$) for   $n\in \Lambda$.
\end{remark}

Now we  use some examples to show how Theorem \ref{MCIeq} can be  flexibly used to derive various sampling formulas for bandlimited signals. For simplicity we shall restrict to the case where $N_2=-N_1=N$.

\begin{example}\label{ex1}
	We first present the most basic example for our results. Let $M=1$, $t_p=\frac{2\pi p}{2N+1}$, $b(n)=1$ . By Theorem \ref{MCIeq},
	we conclude that   $f\in B_\mathbf{N}$ can be recovered from its samples. That is,
	\begin{equation}\label{triinterp}
	f(t)=\frac{1}{2N+1}\sum_{p=0}^{2N} f(t_p)D_N (t-t_p)
	\end{equation}
	where
	\begin{equation*}
	D_N(t)= \begin{cases}
	\frac{\sin(\frac{1}{2}+N)t}{\sin \frac{1}{2}t},&   t\neq 2k \pi,  \\
	2N+1 &  t=2k \pi;
	\end{cases}
	\end{equation*}
	is the $N$-th order Dirichlet kernel. This formula is referred to trigonometric interpolation  for data $f(t_p)$ ($p=0,1,\dots,2N$). It should be stressed that the  interpolation formulas given in \cite{schanze1995sinc,candocia1998comments,dooley2000notes} are mathematically equivalent to  Eq. (\ref{triinterp}).   Note that the expression (\ref{triinterp}) is not numerically stable  at which the function takes $0/0$ form.   Based on the DFT, the author in \cite{candocia1998comments} obtained a numerically stable formulation which is somehow equivalent to interpolation by the FFT \cite{fraser1989interpolation}.  In fact, the DFT or FFT based expressions for trigonometric interpolation can be  subsumed in  Eq. (\ref{DFTexpression}).
\end{example}

\begin{example}\label{ex2}
	Theorem  \ref{MCIeq} permits us to express  $f\in B_{\mathbf{N}}$ from its samples and samples of its  derivatives. Note that
	\begin{equation*}
	f'(t)  = \sum_{n\in I^{\mathbf{N}}} \i n a(n) e^{\i n t},~~  f''(t)  =  \sum_{n\in I^{\mathbf{N}}} - n^2 a(n) e^{\i n t}.
	\end{equation*}
	Let $g_1(t)=f(t),g_2(t)= f'(t),g_3(t)=f''(t) $, then $\H_n$  should be defined by
	\begin{equation*}
	\H_n= \begin{bmatrix}
	1 & \i n  & - n^2 \\
	1 & \i (n+L) & -(n+L)^2\\
	1& \i (n+2L) &-(n+2L)^2
	\end{bmatrix}.
	\end{equation*}
	By straightforward computations,
	\begin{equation*}
	\H_n^{-1}= \begin{bmatrix}
	\frac{2 L^2+3 L n+n^2}{2 L^2} &-\frac{n (2 L+n)}{L^2}& \frac{n (L+n)}{2 L^2} \\
	\frac{\i (3 L+2 n)}{2 L^2} & -\frac{2 \i (L+n)}{L^2} & \frac{\i (L+2 n)}{2 L^2} \\
	-\frac{1}{2L^2} & \frac{1}{L^2} & -\frac{1}{2 L^2}\\
	\end{bmatrix}.
	\end{equation*}
	In order to  satisfy  the assumption that $2N+1$  is divisible by $M=3$, we have to assume that  $N=1 ~~({\rm mod} ~3)$. Suppose that $N=3N_0+1$, then $L=\frac{2N+1}{3}=2N_0+1$. Consequently, we obtain $v_{n,m}(t) $ as follows:
	\begin{align*}
	v_{1,n}(t)=  & \frac{2 L^2+3 L n+n^2}{2 L^2} e^{\i nt} -\frac{n (2 L+n)}{L^2}e^{\i (n+L) t} \\ &{+}\: \frac{n (L+n)}{2 L^2}e^{\i (n+2L) t}, \\
	v_{2,n}(t)= &   \frac{\i (3 L+2 n)}{2 L^2} e^{\i nt}  -\frac{2 \i (L+n)}{L^2} e^{\i (n+L) t} \\ &{+}\:  \frac{\i (L+2 n)}{2 L^2}e^{\i (n+2L) t}, \\
	v_{3,n}(t)=  &    \frac{-1}{2L^2} e^{\i nt} +\frac{1}{L^2}e^{\i (n+L) t} -\frac{1}{2L^2}e^{\i (n+2L) t}.
	\end{align*}
	By using $L=2N_0 + 1$ and  letting $p=0$ in Eq. (\ref{transformym}), we have that $y_1(t)$ equals
	\begin{equation*}
	\frac{ \sin ^3\left(\left({N_0}+\frac{1}{2}\right) t\right) \left({N_0}^2+{N_0}+1-({N_0}+1) {N_0} \cos t\right)}{(2 {N_0}+1)^2\sin ^3\left(\frac{t}{2}\right)}
	\end{equation*}
	and
	\begin{align*}
	y_2(t) &= \sum _{n\in I_1}v_{2,n}(t)=\frac{\sin (t)\sin ^3\left(\left( {N_0}+\frac{1}{2}\right) t\right)}{(2 {N_0}+1)^2 \sin ^3\left(\frac{t}{2}\right) },\\
	y_3(t)&=\sum _{n\in I_1}v_{3,n}(t)=\frac{2  \sin ^3\left(\left({N_0}+\frac{1}{2}\right) t\right)}{(2 {N_0}+1)^2\sin \left(\frac{t}{2}\right)}.
	\end{align*}
	
	Note that for $b_m(n)=(\i n)^{m-1}$ ($1\leq m \leq M$), the determinant of $\H_n$ defined by (\ref{defH}) is a Vandermonde determinant. Therefore $\H_n$  is invertible for all $n\in I_1$. This would  imply that we can further derive   a sampling formula for recovering $f$ from its samples as well as the samples of its first $M-1$ derivatives.
\end{example}

Having considered interpolation for the samples of   original signal and its derivatives. We now continue to show how the samples of Hilbert transform of $f$ can be used to reconstruct $f$ itself. Interestingly, based on  anti-involution property (\ref{Hsquare}),   MCI can be applied  to compute Hilbert transform as well.

\begin{example}\label{ex3}
	Let $M=1$ and $b(n)=-\i\sgn(n)$. Note that $\H_n=b(n)$ does not satisfy the conditions of Theorem \ref{MCIeq} because  it is not   invertible at $n=0$. Nevertheless, By Remark \ref{aequal0}, this inconsistence  can be avoided if $a(0)=0$. Since $b(n)^{-1}=\i \sgn(n)$ for $n\neq 0$, then $y(t)$ equals
	\begin{equation*}
	\sum_{0<\abs{n}\leq N} \i \sgn(n) e^{\i n t }= -2\csc (\frac{t}{2}) \sin(\frac{N t}{2}) \sin\left(\frac{1}{2}(1+N)t\right)
	\end{equation*}
	and the sampling formula
	\begin{equation}\label{samplbyHilberteqn}
	\begin{split}
	f(t)=& \tfrac{-2}{2N+1} \sum_{p=0}^{2N}\mathcal{H}f(t_p) \csc (\tfrac{t-t_p}{2}) \\
	& ~~~~~~~~~~~~{\times} \sin(\tfrac{N (t-t_p)}{2}) \sin\left(\tfrac{1+N}{2}(t-t_p)\right)
	\end{split}
	\end{equation}
	holds if $f\in B_{\mathbf{N}}$ and $a(0)=0$.
	
	Note that   $\mathcal{H}f(t)= \sum_{n\neq 0}(-\i\sgn(n))a_n e^{\i n t}$  satisfies the  preconditions for establishment of  Eq. (\ref{samplbyHilberteqn}) inherently. By substituting $f$ with $\mathcal{H}f$ in (\ref{samplbyHilberteqn}) and using (\ref{Hsquare}), we conclude that the
	Hilbert transform  of $f\in B_{\mathbf{N}}$ can be computed by
	\begin{equation}\label{computeHilbert1}
	\begin{split}
	\mathcal{H}f(t)= &\tfrac{2}{2N+1} \sum_{p=0}^{2N}\left(f(t_p)-a(0)\right) \csc (\tfrac{t-t_p}{2}) \\
	&  ~~~~~~~~~~~~{\times}\sin(\tfrac{N (t-t_p)}{2}) \sin\left(\tfrac{1+N}{2}(t-t_p)\right).
	\end{split}
	\end{equation}
	Unlike   (\ref{samplbyHilberteqn}),  the formula  (\ref{computeHilbert1})  is valid for the case of $a(0)\neq 0$. In fact, if $f\in B_{\mathbf{N}}$, we have
	\begin{equation}\label{a01}
	a(0)= \frac{1}{2 \pi} \int_{\T}f(t)dt = \frac{1}{2N+1}\sum_{p=0}^{2N}f(t_p)
	\end{equation}
	by direct computation.
\end{example}
\begin{remark}
	Given a real-valued $(2N+1)$-point discrete signal $\{x_r(p)\}$, $p=0,1,\dots,2N$.  By replacing $f(t_p)$ with $x_r(p)$ and letting $t=t_j=\tfrac{2\pi j}{2N+1}, j=0,1,\dots,2N$ in  the right hand side of (\ref{computeHilbert1}), we get a discrete signal $x_i(j)=\mathcal{H} f(t_j)$.  The  DFT for $\{z(j)=x_r(j)+\i x_i(j)\}$  is
	\begin{equation*}
	Z(k)= \begin{cases}
	X(0),&  k=0,  \\
	2 X(k), &   1 \leq k\leq N  ,\\
	0, & N+1 \leq k\leq 2N ,
	\end{cases}
	\end{equation*}
	where $X(\cdot)$ is the DFT of $x_r(\cdot)$. That means that $x_i$ is the discrete Hilbert transform of $x_r$ \cite{marple1999computing}.  This fact reveals the relationship  between discrete Hilbert transform and continuous  circular Hilbert transform. For  the  discrete signal   of even number, similar arguments can be made,  we omit the details.
\end{remark}
\begin{example}\label{ex4}
	The analytic signal associated with $f$ is defined by $f_A(t)=f(t)+\i \mathcal{H}f(t)$, it   can be rewritten as
	\begin{equation*}
	f_A(t)=\sum_{n=0}^{N}\widetilde{a_n} e^{\i n t}
	\end{equation*}
	where $\widetilde{a_n}=a_n$ for $n=0$ and $\widetilde{a_n}=2a_n$ for $n\neq0$.
	Applying Theorem \ref{MCIeq} to $f_A(t)$, we have the following.
	If $f\in B_{\mathbf{N}}$,  then
	\begin{equation*}
	f(t)+\i \mathcal{H}f(t)= \frac{1}{1+N}\sum_{p=0}^N  \left(f(t_p)+\i \mathcal{H}f(t_p)\right)y(t-t_p),
	\end{equation*}
	where $t_p=\frac{2\pi p}{N+1}$ and $y(t)=\sum_{n=0}^Ne^{\i n t}= y_r(t)+\i y_i(t)$  with
	\begin{equation*}
	y_r(t) = \csc (\frac{t}{2}) \cos(\frac{N t}{2}) \sin\left(\frac{1}{2}(1+N)t\right)
	\end{equation*}
	and
	\begin{equation*}
	y_i(t) = \csc (\frac{t}{2}) \sin(\frac{N t}{2}) \sin\left(\frac{1}{2}(1+N)t\right).
	\end{equation*}
	Moreover, if $f$ is real-valued, then
	\begin{align}
	f(t) &= \tfrac{1}{1+N}\sum_{p=0}^N f(t_p)y_r(t-t_p) - \mathcal{H}f(t_p)y_i(t-t_p),\label{realvalued2}\\
	\mathcal{H} f(t) &= \tfrac{1}{1+N}\sum_{p=0}^N f(t_p)y_i(t-t_p) + \mathcal{H}f(t_p)y_r(t-t_p). \label{realvalued2H}
	\end{align}
\end{example}

\begin{example}\label{ex5}
	In Example \ref{ex4}, we express $f$ (or $\mathcal{H}f$) by the samples of $f$ and $\mathcal{H}f$. However,  the formulas (\ref{realvalued2}) and (\ref{realvalued2H}) are only valid for the real-valued signals. By Theorem \ref{MCIeq},  it is natural to expect that we can  deduce a new expression of  $f$ (resp. $\mathcal{H}f$)  in terms of the samples of $f$ and $\mathcal{H}f$ by letting $M=2,b_1(n)=1,b_2(n)=-\i\sgn(n)$.
	Rewrite $f\in B_{\mathbf{N}}$ as
	$f(t)=\sum_{n=-N}^{N+1}a_n e^{\i n t }$ with $a_{N+1}=0$.
	Let $M=2, L=N+1$ and $b_1(n)= 1, b_2(n)= -\i \sgn(n)$. Then
	$g_1(t)=f(t)$, $g_2(t) = \mathcal{H}f(t)$ and
	\begin{equation*}
	\H_n= \begin{bmatrix}
	1 & - \i \sgn (n)   \\
	1 & - \i \sgn(n+L)
	\end{bmatrix}.
	\end{equation*}
	It is clear that
	\begin{equation*}
	\H_n^{-1}= \begin{bmatrix}
	\frac{1}{2} &\frac{1}{2}  \\
	- \frac{\i}{2} & \frac{\i}{2}
	\end{bmatrix} ~~\text{for}~~-N\leq n \leq -1,~~ H_0^{-1}= \begin{bmatrix}
	1 & 0 \\
	- \i  &1
	\end{bmatrix}.
	\end{equation*}
	Then by  definition  (\ref{ym}), we get that
	\begin{align*}
	y_1(t)  &  =1 + \sum_{n=-N}^{-1 } \frac{1}{2} \left(e^{\i t (n+N+1)}+e^{\i n t}\right) \\
	& = \frac{\cos ((N+1) t)-\cos (N t)+\cos (t)-1}{2 \cos (t)-2},\\
	y_2(t) & =  e^{\i (N+1) t}-\i -\frac{\i \left(e^{-\i N t}-1\right) \left(e^{\i (N+1) t}-1\right)}{2 \left(e^{\i t}-1\right)}.
	\end{align*}
	Consequently, we obtain
	\begin{equation}\label{complexbyHil2}
	f(t)= \frac{1}{1+N}\sum_{p=0}^N  f(t_p)y_1(t-t_p) + \mathcal{H}f(t_p)y_2(t-t_p)
	\end{equation}
	where $t_p = \frac{2\pi p}{N+1}$.
	Substituting  $f$ with $\mathcal{H}f$ in Eq. (\ref{complexbyHil2}) and applying (\ref{Hsquare}) again, we have that $\mathcal{H}f(t)$ equals
	\begin{equation}\label{complexbyHil2H}
	\frac{1}{1+N}\sum_{p=0}^N  \mathcal{H}f(t_p)y_1(t-t_p) +\left( a(0)- f(t_p)\right)y_2(t-t_p).
	\end{equation}
	The formulas  (\ref{complexbyHil2}) and (\ref{complexbyHil2H}) are applicable to complex-valued signals.
	By the similar arguments to (\ref{a01}), the value of $a(0)$ can be computed by the samples of $f$ and $\mathcal{H}f$. Concretely,
	\begin{equation}\label{a02}
	a(0)=\frac{1}{1+N}\sum_{p=0}^N  f(t_p) + \i \mathcal{H}f(t_p)
	\end{equation}
	holds for $f\in B_{\mathbf{N}}$.
\end{example}

\begin{example}\label{ex6}
	In Example \ref{ex2},  the problem of recovering $f$ from its samples as well as the samples of its derivatives was studied. We are motivated by anti-involution property (\ref{Hsquare}) to consider  the problem of recovering $\mathcal{H}f$ from the   samples  of $f$  along with its derivatives.
	We may use Theorem \ref{MCIeq}  to derive the sampling formula via finding the Fourier multipliers $b_1(n),b_2(n),b_3(n)$ for transforming
	$\mathcal{H} f$ to $f,f',f''$.
	However, from Example  \ref{ex2},  we may obtain  a formula for reconstructing $\mathcal{H}f$ by taking Hilbert transform to $y_m(t-t_p)$ directly. To  achieve this,  by making  use of (\ref{transformym}) and  the linearity of Hilbert transform, it suffices to obtain the Hilbert transform of  $v_{m,n}(t)$ derived in Example \ref{ex2}. By invoking
	\begin{equation*}
	\mathcal{H}\{e^{\i n t}\} =  -\i \sgn(n)e ^{\i  n t },
	\end{equation*}
	it is easy to get $\mathcal{H}\{v_{m,n}\}$ and therefore we can immediately obtain $\mathcal{H}\{y_m(\cdot-t_p)\}$. We omit the explicit expression of $\mathcal{H}\{y_m(\cdot-t_p)\}$ for its lengthiness.
\end{example}

Several  examples have been presented to illustrate how to use  the proposed MCI  flexibly. In these examples, $I^\mathbf{N}$ is assumed to be symmetric to the origin for   simplicity.
For general $I^\mathbf{N}$, the expressions for $y_1(t),y_2(t),\dots,y_M(t)$ could be very  complicated. Moreover the expressions for $y_1(t),y_2(t),\dots,y_M(t)$ (see Examples mentioned earlier), in general, are of $0/0$ form at some points. In fact, we just need to compute $v_{1,n}(t),v_{2,n}(t),\dots,v_{M,n}(t)$ and then use the  numerically stable  expression (\ref{DFTexpression})  for implementation. It should be stressed that the expressions of $v_{1,n}(t),v_{2,n}(t),\dots,v_{M,n}(t)$  would not  change with $I^\mathbf{N}$ and the expression (\ref{DFTexpression}) could be computed cheaply by FFT.

Intensively, we have studied the formulas for approximating $f$ or $\mathcal{H}f$. However, the applications of the proposed MCI go further than that. Generally speaking,  it is applicable to the problem of reconstructing or approximating  $\mathcal{Q}(f)$ from the samples of  $\mathcal{Q}_1(f),\mathcal{Q}_2(f),\dots,\mathcal{Q}_M(f)$, where the operators $\mathcal{Q}$ and $\mathcal{Q}_m$,($1\leq m \leq M$) are determined by  specific needs in practice.

\section{Interpolation consistency and error analysis}\label{S4}

In  practice, the signals (such as chirp signal and Gaussian signal) are   not strictly  bandlimited in general. As we know, if a signal $f(t)$ is not  bandlimited, the reconstructed signal $\widetilde{f}(t)$ given by    any  formula derived in  examples of the previous section is not equal to $f(t)$. In this section,   $f$  is merely assumed to be square integrable on unit circle and is not necessary to be bandlimited.
We define the following approximation operator:
\begin{equation}\label{appxi operator}
\mathcal{T}_\mathbf{N} f(t) : = \frac{1}{L}\sum_{m=1}^{M} \sum_{p=0}^{L-1}g_m(\frac{2\pi p}{L})y_m(t-\frac{2\pi p}{L}).
\end{equation}
There are two natural questions when using multichannel interpolation to reconstruct a non-bandlimited signal.  One is whether  the set of original samples can stay unchanged after reconstruction.  For multichannel interpolation, it is equivalent to examine whether the following equality
\begin{equation}\label{consistency}
\mathcal{T}_\mathbf{N}f*h_m(\tfrac{2\pi k}{L}) = g_m(\tfrac{2\pi k}{L})
\end{equation}
still holds for  non-bandlimited  signal $f$. The other question is how to estimate  the errors that  arise in the practical applications when applying (\ref{appxi operator}) to reconstruct  non-bandlimited signals. In this section, we will address these two issues.

\subsection{Consistency of multichannel interpolation}
In this subsection,  we show that the proposed  multichannel interpolation possesses consistency.
\begin{theorem}
	Let $f(t)$ be a signal defined on $\T$ with finite energy (not  necessary to be bandlimited) and $\mathcal{T}_\mathbf{N}$ be given by (\ref{appxi operator}), then the consistency of multichannel interpolation holds as (\ref{consistency}).
\end{theorem}
\begin{nproof}
	Note that the set of Fourier coefficients of $\mathcal{T}_\mathbf{N}f$ is
	\begin{equation*}
	\frac{1}{L} \sum_{j=1}^{M}\sum_{p=0}^{L-1} g_j(\tfrac{2\pi p}{L})e^{-\i n\frac{2\pi p}{L}} r_j(n), \ n\in\mathbb{Z}.
	\end{equation*}	
	It follows that
	\begin{align}
	& \mathcal{T}_\mathbf{N}f*h_m(\tfrac{2\pi k}{L}) \nonumber\\
	=& \sum_{n\in I^{\mathbf{N}}}\frac{1}{L} \sum_{j=1}^{M}\sum_{p=0}^{L-1} g_j(\tfrac{2\pi p}{L})e^{-\i n\frac{2\pi p}{L}} r_j(n) b_m(n)e^{\i n \frac{2\pi k}{L}}\nonumber\\
	=& \sum_{n\in I_1}\sum_{s=0}^{L-1}\frac{1}{L} \sum_{j=1}^{M}\sum_{p=0}^{L-1} g_j(\tfrac{2\pi p}{L})e^{-\i (n+sL)\frac{2\pi p}{L}} r_j(n+sL) b_m(n+sL)e^{\i (n+sL) \frac{2\pi k}{L}}\nonumber\\
	=& \sum_{n\in I_1}\sum_{s=0}^{L-1}\frac{1}{L} \sum_{j=1}^{M}\sum_{p=0}^{L-1} g_j(\tfrac{2\pi p}{L})e^{-\i n\frac{2\pi p}{L}} r_j(n+sL) b_m(n+sL)e^{\i n \frac{2\pi k}{L}}.\label{consistent1}
	\end{align}
	By the definition of $r_j$, we have
	\begin{equation*}
	\sum_{s=0}^{L-1}r_j(n+sL)b_m(n+sL)=\delta(j-m), \quad \forall n\in I_1.
	\end{equation*}
	Note that $I_1$ consists of  $L$ consecutive integers. We have
	\begin{equation*}
	\sum_{n\in I_1} \frac{1}{L} e^{\i n \frac{2\pi}{L}(k-p)}=\delta(k-p).
	\end{equation*}
	Taking the summation over $s, n, j, p$ in turn, (\ref{consistent1}) reduces to
	$g_m(\tfrac{2\pi k}{L})$. The proof is complete.
\end{nproof}

\subsection{Error analysis}
In this subsection, we  analyze the  error estimate for approximation operator (\ref{appxi operator}).
Let $f_{\tau}(t)=f(t-\tau)$ denote the shifted signal.   Note that
\begin{equation*}
\mathcal{T}_\mathbf{N} f_{\tau}(t)= \frac{1}{L}\sum_{m=1}^{M} \sum_{p=0}^{L-1}g_m(\frac{2\pi p}{L},\tau)y_m(t-\frac{2\pi p}{L})
\end{equation*}
where
\begin{equation}\label{shiftgmsamples}
\begin{split}
g_m(\frac{2\pi p}{L},\tau) = & \frac{1}{2\pi}\int_{\T}f(\xi-\tau)h_m(\frac{2\pi p}{L}-\xi)d\xi \\
= & \sum_n a(n)b_m(n)e^{-\i n \tau}e^{\i n \frac{2\pi p}{L}}.
\end{split}
\end{equation}
It is easy to see that $\mathcal{T}_\mathbf{N} $ is not a shift-invariant operator, that means
$\mathcal{T}_\mathbf{N} f(t-\tau) \neq \mathcal{T}_\mathbf{N} f_{\tau}(t)$ in general. The mean square error of approximation for $f_{\tau}$ is given by
\begin{equation*}
\varsigma(f,\mathbf{N},\tau) = \norm{f_{\tau}-\mathcal{T}_\mathbf{N}  f_{\tau}}_2^2= \frac{1}{2\pi}\int_{\T} \abs{f_{\tau}(t)-\mathcal{T}_\mathbf{N}  f_{\tau}(t)}^2 dt.
\end{equation*}
Note that   the period  of signal $2\pi$ is divisible by  $ \frac{2\pi }{L}$ which is the  spacing of the samples, it follows that $\varsigma(f,\mathbf{N},\tau)$ is $\frac{2\pi }{L} $ periodic in $\tau$.
The time shift $\tau$ may be regarded as the phase difference of $f$ and $f_{\tau}$. In most practical applications, the exact phase of signal is unknown \cite{jacob2002sampling}. Hence, we consider the following averaged error:
\begin{equation*}
\varepsilon(f,\mathbf{N}) = \sqrt{\frac{L}{{2\pi }} \int_{0}^{\frac{2\pi}{L}}\varsigma(f,\mathbf{N},\tau)d \tau}.
\end{equation*}

To state our results, we need to introduce two lemmas.

\begin{lemma}\label{rmbm}
	Let $r_m(n)$ and  $b_m(n)$ be given above.  Then
	$\sum_{m=1}^M r_m(n)b_m(n)=1,$  for all  $n\in I^{\mathbf{N}}$
	and
	$\sum_{m=1}^M r_m(n_1)b_m(n_2)=0,$ for all  $n_1,n_2 \in I^{\mathbf{N}}$ satisfying  $\frac{n_1-n_2}{L}\in \mathbb{Z}\setminus \{0\}$.
\end{lemma}
\begin{nproof}
	This is a  direct consequence 	by comparing the results of   both sides of $\H_n\H_n^{-1}=\mathbf{I}$
	and $\H_n^{-1}\H_n=\mathbf{I}$.
\end{nproof}

\begin{lemma}\label{zmn}
	Suppose that $\{a(n)b_m(n)\}_n\in l^1$ and let
	\begin{equation*}
	z_{mn}(\tau):= \sum_{k}a(n+kL)b_m(n+kL)e^{-\i kL \tau}.
	\end{equation*}
	Then $z_{mn}(\tau)$  is well-defined for every $n\in\mathbb{Z}$ and furthermore
	\begin{equation}\label{lemma11}
	z_{mn}(\tau) = \frac{e^{\i n \tau}}{L}  \sum_{p=0}^{L-1} g_m(\frac{2\pi p}{L},\tau)e^{-\i n \frac{2\pi p}{L}}.
	\end{equation}
\end{lemma}
\begin{nproof}
	Note that  the both sides of   Eq. (\ref{lemma11}) are continuous, to prove Eq. (\ref{lemma11}), it suffices to verify that  the both sides possess  the same Fourier series coefficients. We now need to compute
	\begin{equation*}
	\frac{L}{{2\pi }} \int_{0}^{\frac{2\pi}{L}}  \frac{e^{\i n \tau}}{L}  \sum_{p=0}^{L-1} g_m(\frac{2\pi p}{L},\tau)e^{-\i n \frac{2\pi p}{L}} e^{\i k  {L}\tau} d\tau.
	\end{equation*}
	Under the assumption of $\{a(n)b_m(n)\}_n\in l^1$ ,  the series (\ref{shiftgmsamples}) converges uniformly in $\tau$. Substituting  $g_m(\frac{2\pi p}{L},\tau)$ with the series representation and note that the uniform convergence permits the   interchange of integral and  infinite summation, we get that
	\begin{equation*}
	\begin{split}
	&\tfrac{L}{{2\pi }} \int_{0}^{\frac{2\pi}{L}}  \tfrac{e^{\i n \tau}}{L}  \sum_{p=0}^{L-1} g_m(\frac{2\pi p}{L},\tau)e^{-\i n \frac{2\pi p}{L}} e^{\i k  {L}\tau} d\tau. \\
	= &  \tfrac{L}{{2\pi }} \int_{0}^{\frac{2\pi}{L}}  \tfrac{e^{\i n \tau}}{L}  \sum_{p=0}^{L-1}\sum_l a(l)b_m(l)e^{-\i l \tau}e^{\i l \frac{2\pi p}{L}} e^{-\i n \frac{2\pi p}{L}} e^{\i k  {L}\tau} d\tau\\
	=&  \tfrac{1}{L} \sum_{p=0}^{L-1} \sum_l a(l)b_m(l) e^{\i l \frac{2\pi p}{L}} e^{-\i n \frac{2\pi p}{L}}
	\tfrac{L}{{2\pi }} \int_{0}^{\frac{2\pi}{L}} e^{\i(n+kL-l)\tau}d\tau \\
	=&  \tfrac{1}{L} \sum_{p=0}^{L-1} \sum_l a(l)b_m(l) e^{\i l \frac{2\pi p}{L}} e^{-\i n \frac{2\pi p}{L}} \delta(n+kL-l) \\
	= & \tfrac{1}{L} \sum_{p=0}^{L-1} a(n+kL)b_m(n+kL) e^{\i( n+kL) \frac{2\pi p}{L}} e^{-\i n \frac{2\pi p}{L}}\\
	= & a(n+kL)b_m(n+kL).
	\end{split}
	\end{equation*}
	Here, $\delta(n)=1$ if $n=0$ and $\delta(n)=0$ otherwise.
\end{nproof}

We now use the Parseval's identity to compute $\varsigma(f,\mathbf{N},\tau)$.  By direct computations, we obtain the Fourier   coefficients of $f_{\tau}$ and  $\mathcal{T}_\mathbf{N}  f_{\tau}$ respectively as $  f_{\tau} ~ {\sim}~ a(n)e^{-\i n \tau} $  and
\begin{equation*}
\mathcal{T}_\mathbf{N}  f_{\tau}  ~ {\sim}~\frac{1}{L}\sum_{m=1}^M \sum_{p=0}^{L-1}g_m(\frac{2\pi p}{L},\tau) e^{-\i n \frac{2\pi p}{L}} r_m(n).
\end{equation*}
By Lemma \ref{zmn}, we have $ \mathcal{T}_\mathbf{N}  f_{\tau}  ~ {\sim}~e^{-\i n \tau}\sum_{m=1}^M r_m(n)z_{mn}(\tau)$.
It follows that
\begin{align}
\frac{1}{2\pi} \int_{\T}\abs{f_{\tau}(t) }^2 dt&= \sum_{n\in \mathbb{Z}} \abs{a(n)}^2,\nonumber \\
\frac{1}{2\pi} \int_{\T} \overline{f_{\tau}(t) } \mathcal{T}_\mathbf{N}  f_{\tau}(t)dt & = \sum_{n\in I^{\mathbf{N}}} \overline{a(n)}\sum_{m=1}^M r_m(n)z_{mn}(\tau), \label{ftimeTf}\\
\frac{1}{2\pi} \int_{\T}\abs{\mathcal{T}_\mathbf{N}  f_{\tau}(t) }^2 dt & = \sum_{n\in I^{\mathbf{N}}} \abs{\sum_{m=1}^M r_m(n)z_{mn}(\tau)}^2. \label{TftimeTf}
\end{align}
Integrating both sides of Eq. (\ref{ftimeTf}) with respect to $\tau$, we have that
\begin{equation}\label{average1}
\begin{split}
&\frac{L}{{2\pi }} \int_{0}^{\frac{2\pi}{L}}d\tau \frac{1}{2\pi} \int_{\T} \overline{f_{\tau}(t) } \mathcal{T}_\mathbf{N}  f_{\tau}(t)dt\\
=  &  \sum_{n\in I^{\mathbf{N}}} \overline{a(n)}\sum_{m=1}^M r_m(n) \frac{L}{{2\pi }} \int_{0}^{\frac{2\pi}{L}}z_{mn}(\tau)d\tau\\
=&  \sum_{n\in I^{\mathbf{N}}} \overline{a(n)}\sum_{m=1}^M r_m(n)a(n)b_m(n)\\
=&   \sum_{n\in I^{\mathbf{N}}} \abs{a(n)}^2\sum_{m=1}^M r_m(n)b_m(n)
=\sum_{n\in I^{\mathbf{N}}} \abs{a(n)}^2.
\end{split}
\end{equation}
Here we used the fact that  $\frac{L}{{2\pi }} \int_{0}^{\frac{2\pi}{L}}z_{mn}(\tau)d\tau=a(n)b_m(n)$. The last equality is a consequence of
$\sum_{m=1}^M r_m(n)b_m(n)=1$. With the same arguments to Eq. (\ref{average1}), we have
\begin{equation*}
\frac{L}{{2\pi }} \int_{0}^{\frac{2\pi}{L}}d\tau \frac{1}{2\pi} \int_{\T} {f_{\tau}(t)} \overline{\mathcal{T}_\mathbf{N}  f_{\tau}(t)}dt=\sum_{n\in I^{\mathbf{N}}} \abs{a(n)}^2.
\end{equation*}
By invoking the Parseval's identity again, for $1\leq l,m \leq M$, we get that
\begin{equation*}
\begin{split}
& \frac{L}{{2\pi }} \int_{0}^{\frac{2\pi}{L}}z_{mn}(\tau) \overline{z_{ln} (\tau)}d\tau\\
=  &  \sum_{k\in \mathbb{Z}} a(n+kL) b_m(n+kL) \overline{a(n+kL) } \overline{ b_l(n+kL)}\\
=&  \sum_{k\in \mathbb{Z}} \abs{a(n+kL)}^2 b_m(n+kL) \overline{ b_l(n+kL)}.
\end{split}
\end{equation*}
Therefore by integrating both sides of Eq. (\ref{TftimeTf}) with respect to $\tau$, we get that
\begin{equation}\label{averageTfTf}
\begin{split}
&\int_{0}^{\frac{2\pi}{L}}d\tau \frac{1}{2\pi} \int_{\T}\abs{\mathcal{T}_\mathbf{N}  f_{\tau}(t) }^2 dt  \\
= & \int_{0}^{\frac{2\pi}{L}}d\tau  \sum_{n\in I^{\mathbf{N}}}  \sum_{m=1}^M \sum_{l=1}^M
r_m(n) \overline{r_l(n)} z_{mn}(\tau) \overline{z_{ln} (\tau)}\\
=&  \sum_{n\in I^{\mathbf{N}}}  \sum_{m=1}^M \sum_{l=1}^M r_m(n) \overline{r_l(n)} \sum_{k\in \mathbb{Z}} \abs{a(n+kL)}^2\\
& ~~~~~~~~~~~~~~~~~{\times} b_m(n+kL) \overline{ b_l(n+kL)}\\
=& \sum_{k\in \mathbb{Z}}  \sum_{n\in I^{\mathbf{N}}} \abs{a(n+kL)}^2 \abs{\sum_{m=1}^M r_m(n)b_m(n+kL)}^2.
\end{split}
\end{equation}
By a  change of variables  and note that $J_k= \bigcup_{l=k+1}^{M+k}I_l$, Eq.   (\ref{averageTfTf}) can be rewritten as
\begin{equation*}
\sum_{k\in \mathbb{Z}}\sum_{n\in J_k} \abs{a(n)}^2 \abs{\sum_{m=1}^M r_m(n-kL)b_m(n)}^2.
\end{equation*}
From Lemma \ref{rmbm}, we have that
\begin{equation*}
\sum_{n\in{J_0}}\abs{a(n)}^2 \abs{\sum_{m=1}^M r_m(n)b_m(n)}^2 = \sum_{n\in I^{\mathbf{N}}} \abs{a(n)}^2
\end{equation*}
and
\begin{equation*}
\sum_{k\neq 0 } \sum_{n\in{J_k\cap I^{\mathbf{N}}}}\abs{a(n)}^2 \abs{\sum_{m=1}^M r_m(n-kL)b_m(n)}^2 =0.
\end{equation*}
Then Eq. (\ref{averageTfTf}) reduces to
\begin{equation*}
\sum_{n\in I^{\mathbf{N}}} \abs{a(n)}^2 +  \sum_{k\neq 0 }\sum_{n\in{J_k- I^{\mathbf{N}}}}\abs{a(n)}^2 \abs{\sum_{m=1}^M r_m(n-kL)b_m(n)}^2.
\end{equation*}
Rearranging the terms,  the second part of the above equation    becomes
\begin{equation*}
\sum_{k\notin\{1,2,\dots,M\}}\sum_{n\in  I_k}\abs{a(n)}^2 \sum_{l=1}^{M}\abs{\sum_{m=1}^M r_m(n+(l-k)L)b_m(n)}^2.
\end{equation*}
Combining the computations above,    $\frac{L}{{2\pi }} \int_{0}^{\frac{2\pi}{L}}\varsigma(f,\mathbf{N},\tau)d \tau$  (namely  $\varepsilon(f,\mathbf{N})^2$) can be expressed as Eq. (\ref{averageerror}). Here,
\begin{align}
\varepsilon(f,\mathbf{N})^2&=     \sum_{n\in \mathbb{Z}} \abs{a(n)}^2 -2 \sum_{n\in I^{\mathbf{N}}} \abs{a(n)}^2+\sum_{n\in I^{\mathbf{N}}} \abs{a(n)}^2 \nonumber \\
&~~~~+ \sum_{k\notin\{1,2,\dots,M\}}\sum_{n\in  I_k}\abs{a(n)}^2 \sum_{l=1}^{M}\abs{\sum_{m=1}^M r_m(n+(l-k)L)b_m(n)}^2\nonumber \\
&=  \sum_{n\notin I^{\mathbf{N}}} \abs{a(n)}^2+ \sum_{k\notin\{1,2,\dots,M\}}\sum_{n\in  I_k}\abs{a(n)}^2 \sum_{l=1}^{M}\abs{\sum_{m=1}^M r_m(n+(l-k)L)b_m(n)}^2  \label{averageerror} \\
&\leq    \sum_{n\notin I^{\mathbf{N}}} \abs{a(n)}^2+ \sum_{n\notin I^{\mathbf{N}}}\abs{a(n)}^2 \left( \sum_{m=1}^M \abs{b_m(n)}^2\right)
\sum_{l=1}^{M}\left(\sum_{k=1}^M\abs{\Omega_k(L)}^2\right)\nonumber \\
&=  \sum_{n\notin I^{\mathbf{N}}} \abs{a(n)}^2+ \sum_{n\notin I^{\mathbf{N}}}\sum_{m=1}^M \abs{a(n)b_m(n)}^2
M\left(\sum_{k=1}^M\abs{\Omega_k(L)}^2\right).  \label{errorestimate}
\end{align}
Accordingly, we have the following theorem.
\begin{theorem}\label{error}
	Let $f\in L^2(\T)$,  then the expression of $\varepsilon(f,\mathbf{N})$ is given by square root of Eq. (\ref{averageerror}).
\end{theorem}

Let $\mu (I^{\mathbf{N}})=LM$ denotes the total number of elements of $I^{\mathbf{N}}$.  For fixed $M$ and $b_1,b_2,\dots,b_M$, roughly speaking, $\varepsilon(f,\mathbf{N})$ tends to $0$ as
$\mu (I^{\mathbf{N}})\to \infty$ (or $L\to \infty$) if the rate of $ \abs{a(n)b_m(n)}\rightarrow 0 $  is sufficiently fast as $\abs{n}\to \infty$. There are no unified conditions to ensure  $\varepsilon(f,\mathbf{N})\to 0$ as $\mu (I^{\mathbf{N}})\to \infty$, it should be concretely analyzed for the different situations. This is due to the fact that   $r_m(n)$ is constructed from $\H_n^{-1}$ which depends on $\mu (I^{\mathbf{N}})$ (or $L$).  Nevertheless,  a rough estimation can be given  for  Eq. (\ref{averageerror}).
For $1\leq m \leq M$, let
\begin{equation*}\label{boundforrm}
\Omega_m(b_1,b_2,\dots,b_M,\mu (I^{\mathbf{N}})) =  \sup_{i\in I^{\mathbf{N}}} \abs{r_m(i)}^2.
\end{equation*}
Keeping in mind that $\Omega_m$ is dependent on $b_1,b_2,\dots,b_M$,   we may rewrite the above equation as
\begin{equation*}
\Omega_m(L) =  \sup_{i\in I^{\mathbf{N}}} \abs{r_m(i)}^2
\end{equation*}
for simplicity. Applying Cauchy-Schwarz inequality to Eq. (\ref{averageerror}), it follows that the value of $\varepsilon(f,\mathbf{N})^2$ is bounded by Eq. (\ref{errorestimate}).

Suppose that there exists a constant $C_1>0$ such that
\begin{equation*}
\sum_{m=1}^M\abs{\Omega_m(L)}^2 \leq C_1 L^\alpha
\end{equation*}
for all large $L$. Then
$\varepsilon(f,\mathbf{N})\to 0$  as $L\to \infty$
if there exists  $C_2>0$  and $\beta< \min\{-1,-\alpha-1\}$ such that
\begin{equation*}
\abs{a(n)b_m(n)}^2 \leq  C_2 \abs{n}^\beta,~~ 1 \leq m  \leq   M
\end{equation*}
for all large $\abs{n}$.
By invoking Eq. (\ref{defrm}), we have
\begin{equation*}
\Omega_m(L) =  \sup_{1\leq j \leq M,i\in I_1} \abs{q_{mj}(i)}^2.
\end{equation*}
For Example \ref{ex2}, note that the factor    $L^2$   in the denominators of $\H_n^{-1}$, it is not difficult to see that $\Omega_1(L)$ is bounded with respect to $L$, and $\Omega_2(L),\Omega_3(L)$ tend to $0$ as $L\to \infty$. Therefore there exists a constant $C_1>0$ such that
$\sum_{m=1}^3\abs{\Omega_m(L)}^2 \leq C_1$
for all $L$.  For Example \ref{ex5}, $\Omega_1(L),\Omega_2(L)$ are independent on $L$, it is obvious that $\sum_{m=1}^2\abs{\Omega_m(L)}^2$ is bounded with respect to $L$.

If $f\in B_{\mathbf{N}}$, the proposed MCI (\ref{drictexpression}) can perfectly reconstruct $f$ by Theorem \ref{MCIeq}.  This fact is also reflected in Eq. (\ref{averageerror}), namely $\varepsilon(f,\mathbf{N})=0$ if $a(n)=0$ for all $n\notin I^{\mathbf{N}}$. It means that if $f\in B_{\mathbf{N}}$, whatever $M$ and $b_1,b_2,\dots,b_M$ we select,  the approximation operator $\mathcal{T}_\mathbf{N} f$  defined by (\ref{appxi operator}) will come into being the same result provided that the total samples used in Eq. (\ref{appxi operator})  is equal to  $\mu (I^{\mathbf{N}})$.
In general, however, for fixed $\mu (I^{\mathbf{N}})$, it follows from Eq. (\ref{averageerror}) that different  $M$ and $b_1,b_2,\dots,b_M$ for $\mathcal{T}_\mathbf{N} f$   may lead to different results if $f\notin B_{\mathbf{N}}$. We can also see this from the numerical examples in the next section.

\section{Numerical examples and applications}\label{S5}

\subsection{ Numerical examples}\label{S51}

\begin{figure*}[!ht]
	\centering
	\includegraphics[width=5in]{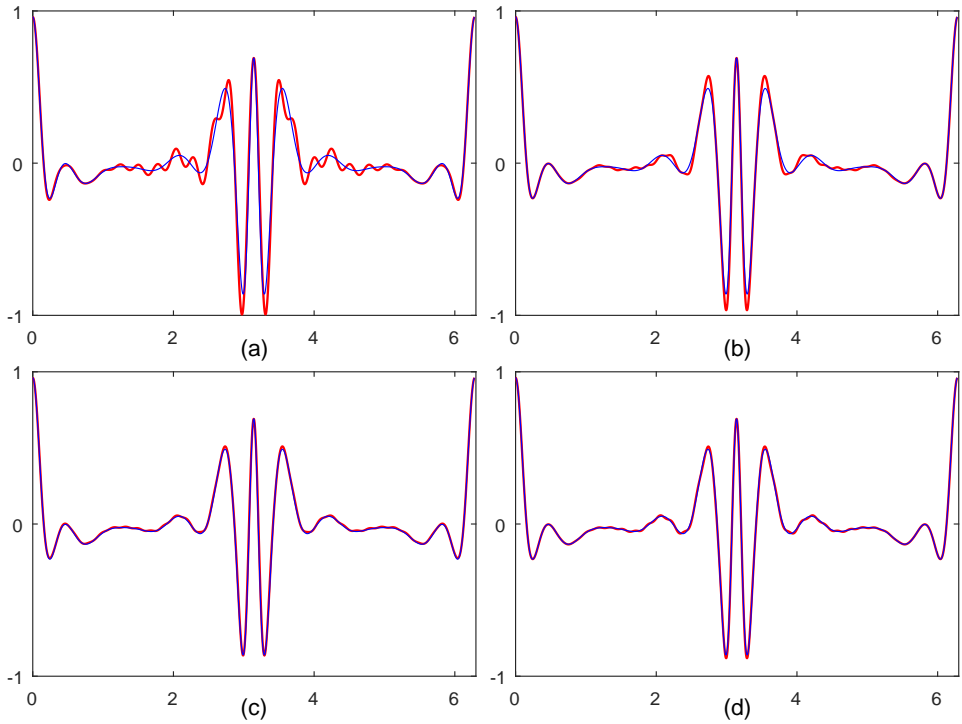}\\
	\caption{ The blue line is the actual $f$.  The red lines are approximated results for $f$ by using  (a) $48$ samples of $f$; (b) $24$ samples of $f$, $24$ samples of $f'$, $24$ samples of $f''$; (c) $36$ samples of $f$, $36$ samples of $\mathcal{H}f$; (d) $72$ samples of $f$. }\label{figapproxf}
\end{figure*}

\begin{figure*}[!ht]
	\centering
	\includegraphics[width=5in]{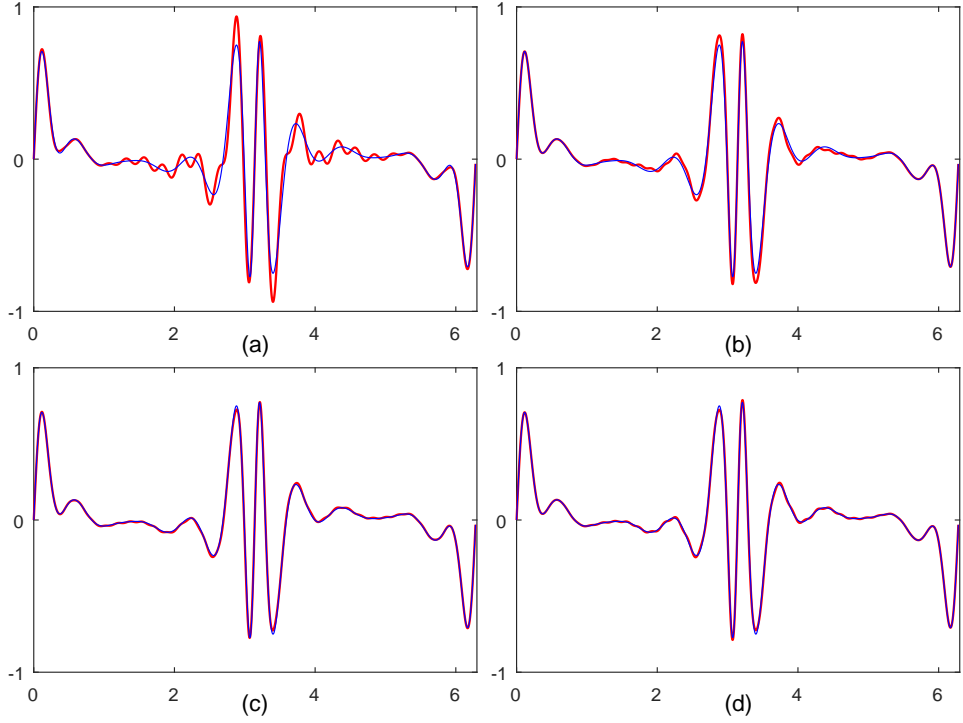}\\
	\caption{ The blue line is the actual $\mathcal{H}f$.  The red lines are approximated results for $\mathcal{H}f$ by using  (a) $48$ samples of $f$; (b) $24$ samples of $f$, $24$ samples of $f'$, $24$ samples of $f''$; (c) $36$ samples of $f$, $36$ samples of $\mathcal{H}f$; (d) $72$ samples of $f$. }\label{figapproxHf}
\end{figure*}

In this part, we shall demonstrate the effectiveness of  $\mathcal{T}_{\mathbf{N}}$ for approximating signals experimentally. We compare the results by using the different formulas derived in Section 3 to reconstruct  $f$ (or $\mathcal{H}f$). Let
\begin{equation*}
\phi(z) = \frac{0.08 z^2+0.06z^{10}}{(1.3-z)(1.5-z)} + \frac{0.05z^3+0.09z^{10}}{(1.2+z)(1.3+z)}.
\end{equation*}
From the theory of Hardy space, the imaginary part of $\phi(e^{\i t })$ is the Hilbert transform of its real part. In the following we select  $f(t)= \Re[\phi(e^{\i t })] $, then its Hilbert transform  is $\mathcal{H}f(t)=\Im[\phi(e^{\i t })]$.

\begin{table}[!ht]
	\caption{Approximated results by using different types of data and different total samples.}\label{approximatedtable}
	\centering
	\begin{tabular}{p{0.6cm}  p{0.32cm} p{0.30cm} p{0.29cm}  p{0.28cm}   l l }
		\hline
		$\mu(I^{\mathbf{N}})$ & ~$f$ & $ \mathcal{H}f~ $  & $ f' $ & $ f'' $ &~~~~~~$\delta_1$ &
		~~~~~~$\delta_2$   \\
		\hline
		$16$  & $16$ &~$0$ & $0$ & $0$ & $0.1482 \times 10^1$ &$ 0.1393 \times 10^1$ \\
		$24$  & $24$ & ~$0$ & $0$ & $0$ & $0.1067 \times 10^1$ &$ 0.1055 \times 10^1$   \\
		$32$  & $16$ & ~$16$ & $0$ & $0$ & $0.9064 \times 10^0$ &$ 0.7532 \times 10^0$   \\
		$32$  & $32$ &~$0$ & $0$ & $0$ & $0.6665\times 10^0$ &$ 0.6653 \times 10^0$   \\
		$48$  & $16$ & ~$0$ & $16$ & $16$ & $0.9066\times 10^0$ &$ 0.8955 \times 10^0$  \\
		$48$  & $24$ & ~$24$ & $0$ & $0$ & $0.2861\times 10^0$ &$ 0.2400\times 10^0$ \\
		$48$  & $48$ &~$0$ & $0$ & $0$ & $0.2126\times 10^0$ &$ 0.2126\times 10^0$ \\
		$72$  & $24$ &~$0$ & $24$ & $24$ & $0.9973\times 10^{-1}$ &$ 0.9947\times 10^{-1}$ \\
		$72$  & $36$ &~$36$ & $0$ & $0$ & $0.3802\times 10^{-1}$ &$ 0.3233\times 10^{-1}$ \\
		$72$  & $72$ & ~$0$ & $0$ & $0$ & $0.2905\times 10^{-1}$ &$ 0.2905\times 10^{-1}$ \\
		$96$  & $32$ & ~$0$ & $32$ & $32$ & $0.1130\times 10^{-1}$ &$ 0.1129\times 10^{-1}$ \\
		$96$  & $48$ & ~$48$ & $0$ & $0$ & $0.4527\times 10^{-2}$ &$ 0.3836\times 10^{-2}$ \\
		$96$  & $96$ &~$0$ & $0$ & $0$ & $0.3494\times 10^{-2}$ &$ 0.3494\times 10^{-2}$ \\
		$108$  & $36$ & ~$0$ & $36$ & $36$ & $0.3803\times 10^{-2}$ &$ 0.3802\times 10^{-2}$ \\
		$108$  & $54$ & ~$54$ & $0$ & $0$ & $0.1537\times 10^{-2}$ &$ 0.1315\times 10^{-2}$ \\
		$108$  & $108$ &~$0$ & $0$ & $0$ & $0.1189\times 10^{-2}$ &$ 0.1189\times 10^{-2}$ \\ \hline
	\end{tabular}
\end{table}

The relative mean square error (RMSE) for approximating $f$ is defined by
\begin{equation*}
\begin{split}
\delta_1 = & \left. \left( \sum_{p=0}^{2047}\abs{f(t_p)-\mathcal{T}_{\mathbf{N}}f(t_p)}^2\right)^{\frac{1}{2}} \middle / \left(\sum_{p=0}^{2047}\abs{f(t_p)}^2\right)^{\frac{1}{2}} \right. \\
\approx &
\left. \left( \int_{\T}\abs{f(t)-\mathcal{T}_{\mathbf{N}}f(t)}^2dt\right)^{\frac{1}{2}} \middle /\left( \int_{\T}\abs{f(t)}^2dt\right)^{\frac{1}{2}} \right.
\end{split}
\end{equation*}
where $t_p=\frac{2\pi p}{2048}$. Similarly, the RMSE for approximating $\mathcal{H}f$ is given by
\begin{equation*}
\delta_2 = \left. \left( \sum_{p=0}^{2047}\abs{\mathcal{H}f(t_p)-\mathcal{T}_{\mathbf{N}}\mathcal{H}f(t_p)}^2\right)^{\frac{1}{2}} \middle / \left(\sum_{p=0}^{2047}\abs{\mathcal{H}f(t_p)}^2\right)^{\frac{1}{2}} \right..
\end{equation*}
All algorithms of experiments are based on FFT representation (\ref{DFTexpression}) and all codes are programmed in Matlab R2016b.   It is easy to see that the computational complexity for computing $N_{out}$ number of  functional values of  $f(t)$ by  Eq. (\ref{DFTexpression}) is
\begin{equation}\label{complexityGSE}
\mathcal{O} (N_{out} M  L  \log L).
\end{equation}
If $N_{out}$ is a multiple of $L$, namely $N_{out} = C L$, then the computational complexity is reduced to
$\mathcal{O} ((C-1)M  L^2  \log L)$. Note that $M$ is the number of data types, meaning that it is usually a small integer.

The approximated results by using different types of data and different total samples are listed in Table \ref{approximatedtable}.
The first column $\mu(I^{\mathbf{N}})$ is the total number of samples used in each experiment. The column 2 to 5 are respectively the number of samples of $f$, $\mathcal{H}f$, $f'$, $f''$ used in each experiment.
The last two columns are  RMSEs  for approximating $f$ and $ \mathcal{H}f$ respectively. It can be seen that the  experimentally obtained RMSE $\delta_1$ and $\delta_2$  tend to $0$ as $\mu(I^\mathbf{N})$ goes to infinity.
Observe that $\delta_1$ and $\delta_2$ are nearly equal in each row of Table \ref{approximatedtable}.  This is  consistent with  the theoretical prediction, since the Fourier coefficients of $f$ and $\mathcal{H}f$ possess the same absolute value for all $n\in \mathbb{Z} \setminus\{0\}$.
For fixed $\mu(I^\mathbf{N})$,  approximating $f$ (or $\mathcal{H}f$)  from the samples of $f$ performs slightly better than approximating $f$ from the samples of $f$ along with  its Hilbert transform,  and   approximating $f$ from the samples of $f$ along with  its first two order  derivatives has the worst  performance, as compared with the other two approximations. This result is in agreement with the theoretical error estimation (\ref{averageerror}).
Nevertheless, if the same total samples are used to approximate $f$ (or $\mathcal{H}f$), the fluctuations  of RMSEs caused by different types of data  are not significant. We can see this, for example, from the last three rows of Table \ref{approximatedtable}, the RMSEs of approximating from different types of data but with same total samples are in the same order. Generally speaking, the total number of samples  $\mu(I^\mathbf{N})$ is   more crucial than the  types of samples in approximating signals. Figure \ref{figapproxf} and \ref{figapproxHf} may provide an   intuitive reflection for this fact.   It can be seen from both figures that   the rank of    performance in approximation for four experiments is $(a)<(b)<(c)\approx (d)$, where $(a),(b),(c),(d)$ correspond to the experiments of row $7$ to $10$   in Table \ref{approximatedtable}.
It is worth noting that given $\mu(I^\mathbf{N})=LM$ samples associated with $f$, it is flexible to select $\mathbf{N}=\{N_1,N_2\}$  in $\mathcal{T}_\mathbf{N}$ for approximations. If there is some priori information about the frequency of $f$, we may select $\mathbf{N}$ such that the energy of Fourier coefficients of $f$ is concentrated on $I^\mathbf{N}$. Then we can get an optimal approximation which minimizes the RMSE.

\begin{figure*}[!t]
	\centering
	\includegraphics[width=5in]{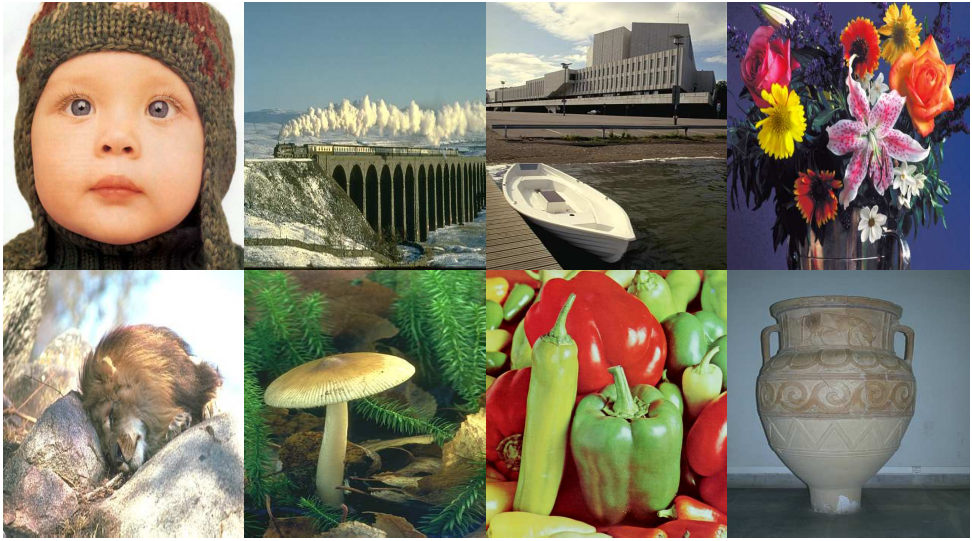}\\
	\caption{The test images. From left to right and top to bottom: \emph{Baby, Bridge, Building, Flowers, Lion, Mushroom, Peppers, Pot.}}\label{originalimage}
\end{figure*}

\subsection{Application to single image super-resolution}

In this part, we apply  MCI to single image super-resolution reconstruction (SISR). 
Our  SISR algorithm is based on Example \ref{ex2}. We first compute interpolation result for each row of image, and then apply interpolation for each column by using the same operations. The derivatives of image are computed   by three-point centered-difference formula. To evaluate the performance of our method, we conduct  experiments  to compare the proposed technique with the  SISR methods aforementioned in the introduction part, including
\begin{itemize}
	\item a classical interpolation-based method (Lanczos);
	\item two improved interpolation-based methods (MCI and GSE \cite{wei2016generalized}), which take image derivative into consideration;
	\item a two-scale method called displacement field \cite{wang2014fast} (FD), which utilizes both interpolation and reconstruction techniques;
	\item  a reconstruction-based method (CNRS \cite{chang2018single}) which takes also advantages of learning techniques;
	\item a improved FD method (MCIFD) by introducing  MCI technique at the interpolation step for FD method. 
\end{itemize}
We note that the interpolation-based methods  are frequently incorporated into other methods  to improve HR results. Thus  we combine our MCI method  with FD method to  verify whether MCI can improve the original FD's performance. We will see from the experimental results that the answer is affirmative.

All the experiments are carried out on an Inter(R) Core (TM) i5-3470s (2.9GHz) PC under the MATLAB R2016b programming environment.
The  SISR results are evaluated by   image quality assessment metrics peak signal to noise ratio (PSNR),
structure similarity (SSIM) index \cite{wang2004image}, feature similarity (FSIM) index \cite{zhang2011fsim} and correlation coefficient (CC) \cite{li2014image} index. The SISR methods are only used to reconstruct the luminance channel since the human visual system is sensitive to the luminance component. We select eight   images  from \emph{Set5} \cite{bevilacqua2012low}, \emph{Set14} \cite{zeyde2010single} and \emph{BSD100} \cite{timofte2014a+} with  different scenes (see Figure \ref{originalimage})  for testing and  the scheme of  experiments is as follows:
\begin{enumerate}
	\item The test images are first blurred by a Gaussian kernel with size of $5\times 5$ and standard deviation of $1.0$, and then downsampled by factor 3.
	\item   Generate high-resolution (HR) images ($\times 3$) from each downsampled image by   different algorithms.
	\item   Compute  PSNR, SSIM, FSIM and CC to evaluate the quality of   generated HR images.
\end{enumerate}

\begin{table}[!t]
	\caption{The PSNR, SSIM, FSIM, and CC results by different methods ($\times 3$). (Bold: the best; underline: the second best).}\label{SISR comparasion}  \vspace{2pt}	\centering \small
	\begin{tabular}{l l l l l l l l}
		\hline
		Image/Method&  & MCI & MCIFD &FD & CRNS & GSE & Lanczos  \\
		\hline
		Baby	&PSNR  & \textbf{33.18} &31.86  &29.00 & \underline{33.13} & 32.85 &29.62\\
		$(510\times 510)$	& SSIM &  \underline{0.8957}& 0.8749 & 0.8481 & \textbf{0.9081} &0.8896&0.8479  \\
		&FSIM  & \underline{0.9770} &0.9578 & 0.9263 &  \textbf{0.9784}& 0.9719& 0.9393\\
		&CC  &\textbf{0.9958}  & 0.9943 & 0.9890 & \textbf{0.9958} &  \underline{0.9955}&0.9904\\
		
		Bridge & PSNR & \textbf{24.18} & 23.32 & 22.76 & 23.71 & \underline{24.09} &23.02\\
		$(480\times 321)$	& SSIM & \underline{0.7485} & 0.7163 & 0.6943 & \textbf{0.7798} & 0.7442 &0.6975\\
		& FSIM & \underline{0.7719} & 0.7152 & 0.6927 & \textbf{0.8158} & 0.7618 &0.7392\\
		& CC & \textbf{0.9385} & 0.9243 & 0.9136 & 0.9335 & \underline{0.9372}&0.9185\\
		
		Building & PSNR & \underline{25.42} & 24.97 & 23.94 & \textbf{25.52} & 25.28 &23.97\\
		$(321\times 480)$	& SSIM & \underline{0.7197} & 0.7063& 0.6765 & \textbf{0.7580} & 0.7131 &0.6607\\
		& FSIM & \underline{0.8139} & 0.7969 & 0.7678 & \textbf{0.8391} & 0.8050&0.7815 \\
		& CC & \underline{0.9751} & 0.9723 & 0.9648 & \textbf{0.9760} & 0.9743 &0.9650\\
		
		Flowers&PSNR & \textbf{26.96} & 26.03 & 24.32 & \underline{26.85} & 26.71 &24.46\\
		$(498\times 360)$	&SSIM& \underline{0.8004} & 0.7694 & 0.7286 & \textbf{0.8321} & 0.7912& 0.7204\\
		&FSIM	& \underline{0.8455} & 0.8193 &0.7825 & \textbf{0.8713} & 0.8394&0.8027 \\
		&CC& \underline{0.9717} & 0.9646 & 0.9469 & \textbf{0.9726} & 0.9700 &0.9486\\
		
		Lion & PSNR & \textbf{26.43} & 25.54 & 24.63 & 25.67 & \underline{26.24} &24.63\\
		$(480\times 321)$	& SSIM & \underline{0.7287} &0.6833 &0.6553   & \textbf{0.7612}&  0.7192 &0.6481\\
		& FSIM & \underline{0.8220} &0.7680& 0.7480  & \textbf{0.8663}& 0.8109&0.7893\\
		& CC & \textbf{0.9682} & 0.9606 & 0.9514& 0.9631&  \underline{0.9668} &0.9513\\
		
		Mushroom & PSNR & \textbf{28.33} & 27.56 & 26.38 & 27.76& \underline{28.11}&26.32\\
		$(321\times 480)$	& SSIM & \underline{0.7652} & 0.7231 &  0.6973& \textbf{0.7921} & 0.7543&0.6786 \\
		& FSIM & \underline{0.8365} & 0.7985 & 0.7749 & \textbf{0.8658}& 0.8263 &0.7973\\
		& CC & \textbf{0.9727} & 0.9673 & 0.9568 &0.9700  &\underline{0.9714}  &0.9561\\
		
		Peppers & PSNR & \underline{31.88} & 31.10 & 27.82 & \textbf{32.14} & 31.60&28.33 \\
		$(510\times 510)$	& SSIM & \underline{0.8642} & 0.8565 &0.8255 &\textbf{0.8752}  & 0.8610& 0.8281\\
		& FSIM & \textbf{0.9764} & 0.9635 & 0.9310& \underline{0.9752} &  0.9723&0.9364\\
		& CC & \underline{0.9901} &0.9879 & 0.9748& \textbf{0.9910} & 0.9895& 0.9774\\
		
		Pot & PSNR &\textbf{33.73}  &33.38  & 32.22 &  \underline{33.60} & 33.52&32.20 \\
		$(321\times 480)$	& SSIM & \underline{0.8591} & 0.8550 & 0.8395 & \textbf{0.8713} &0.8563 &0.8300\\
		& FSIM & \underline{0.8801} &0.8683 & 0.8528 & \textbf{0.8986}&0.8760 &0.8610 \\
		& CC & \textbf{0.9797} & 0.9780 & 0.9712 &\underline{0.9794} & 0.9787&0.9710 \\
		\hline
	\end{tabular}
\end{table}

\begin{figure*}[!t]
	\centering
	\includegraphics[width=6.2in]{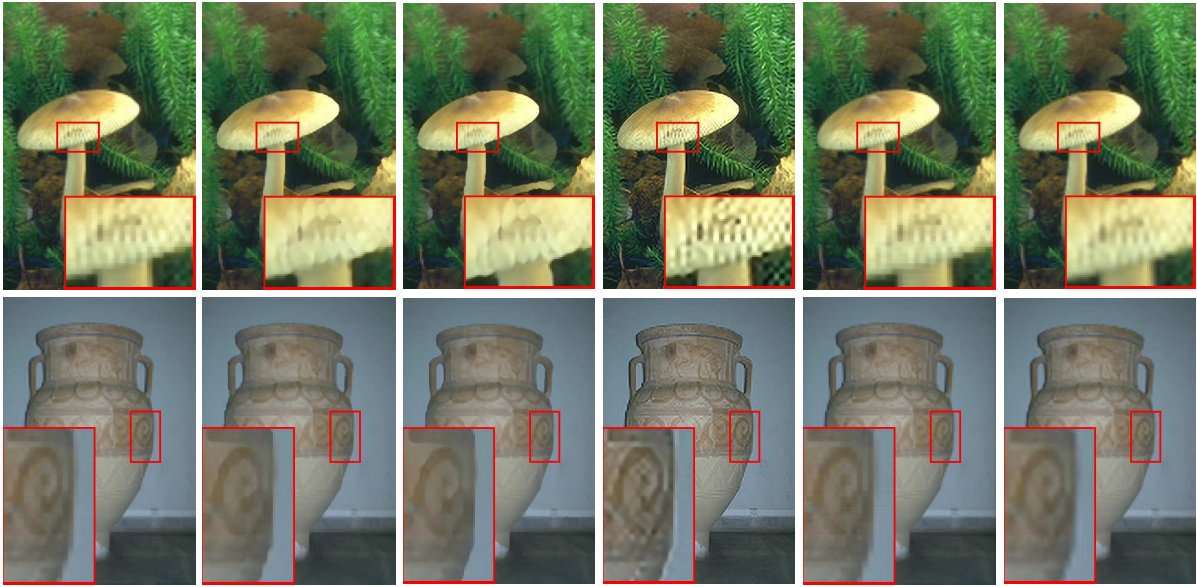}\\
	\caption{SISR results ($\times 3$) of \emph{Mushroom} (top) and \emph{Pot} (bottom)  given by different methods. From left to right: MCI, MCIFD, FD, CRNS, GSE, Lanczos.}\label{reconsimagesnew}
\end{figure*}


\begin{table}[!t]
	\caption{Average  PSNR, SSIM, FSIM, and CC results on datasets by different methods ($\times 3$).  (Bold: the best; underline: the second best).}\label{averageQM}  \vspace{2pt}	\centering \small
	\begin{tabular}{l l l l l l l l}
		\hline
		Metric/Method & & MCI &MCIFD&FD &CRNS & GSE & Lanczos  \\
		\hline
		\emph{Set5}&	PSNR&	\underline{29.87}&	28.60&	26.32&	\textbf{30.08}&	29.51&	26.69\\
		&SSIM&	\underline{0.8613}&	0.8463&	0.8046&	\textbf{0.8861}&	0.8547&	0.8047 \\
		&FSIM&	\underline{0.8990}&	0.8878&	0.8464&	\textbf{0.9172}&	0.8930&	0.8619 \\
		&CC&	\underline{0.9839}&	0.9787&	0.9638&	\textbf{0.9872}&	0.9825&	0.9668  \\
		\emph{Set14}&	PSNR&	\textbf{27.30}&	26.32&	24.71&	\underline{27.15}&	26.31&	24.89 \\
		&SSIM&	\underline{0.7724}&	0.7420&	0.7079&	\textbf{0.8034}&	0.7412&	0.7026  \\
		&FSIM&	\underline{0.9108}&	0.8835&	0.8504&	\textbf{0.9217}&	0.8920&	0.8611 \\
		&CC	&\textbf{0.9650}&	0.9557&	0.9399&	\underline{0.9644}&	0.9480&	0.9423\\
		\emph{BSD100}&	PSNR&	\textbf{27.07}&	26.30&	25.19&	26.67&	\underline{26.92}&	25.25\\
		& SSIM&	\underline{0.7395}	&0.7061	&0.6795&	\textbf{0.7709}&	0.7333&	0.6720\\
		& FSIM	&\underline{0.8318}&	0.7946&	0.7723&	\textbf{0.8671}&	0.8239&	0.8000\\
		& CC &	\textbf{0.9501}&	0.9393&	0.9257&	0.9463&	\underline{0.9486}&	0.9261\\
		\hline
	\end{tabular}
\end{table}

\begin{table}[!t]
	\caption{Average reconstruction time on test images by different methods 
		($\times 3$).}\label{averagetime}  \vspace{2pt}	\centering \small
	\begin{tabular}{l l l l l l l }
		\hline
		Image size/Method  & MCI&MCIFD&FD & CRNS & GSE & Lanczos  \\
		\hline
		$480\times 321$  & 2.61 s &3.14 s  &$<$  1 s& 871.85 s & 15.25 s & $<$ 1 s \\
		$498\times 360$  &  2.85 s& 3.46 s & $<$  1 s& 1023.88 s &19.87 s&$<$  1 s  \\
		$510\times 510$  & 3.75 s & 4.56 s  & $<$  1 s&  1489.39 s&25.61 s& $<$  1 s \\ 		
		\hline
	\end{tabular}
\end{table}



In most real applications, the parameters of image degradation and downsampling are unknown. Thus a perturbation should be added to the parameters in  upsampling   step for getting close to real application scenes. In  our  upsampling   step,   the   window size and the standard deviation are modified as $7\times7$ and $1.4$ respectively. The PSNR, SSIM, FSIM and CC measurements for SISR results are listed in Table \ref{SISR comparasion} and the reconstructed images \emph{Mushroom} and \emph{Pot} are shown in Figure \ref{reconsimagesnew}. Visibly, CNRS produces
high quality HR images with good sharpness and fantastic details. MCI also provides fine HR images,
though not as good as  CNRS's results, but it achieves an improvement over GSE and Lanczos algorithms. FD is a sharpness preserving interpolation technology which adjusts the bicubic upsampling method  by using displacement field. The HR images produced by FD and MCIFD possess relatively clear  edges.

In order to evaluate the performance of  different algorithms more comprehensively, we   further carry out experiments on all of the images in datasets  \emph{Set5}, \emph{Set14} and \emph{BSD100}. The average PSNR, SSIM, FSIM and CC values   in each popular dataset are listed in Table \ref{averageQM}. 
The quantitative results show that CRNS outperforms  the others in SSIM and FSIM significantly. MCI has the second best performance in a quantitative manner and its computational burden is  low  (see Table \ref{averagetime}). The original FD can achieve a evident improvement if we replace bicubic by MCI at the interpolation step. Although FD can effectively restore sharpened edges but it also introduces some over-flat regions. This     brings about a consequence that  MCIFD is slightly inferior to MCI quantitatively.   Since MCI has good high frequency response, it can produce  HR images  with fine detail. Hence MCIFD  achieves a good trade-off between sharp edge preservation and small detail recovery.

\begin{figure*}[!t]
	\centering
	\includegraphics[width=6.3in]{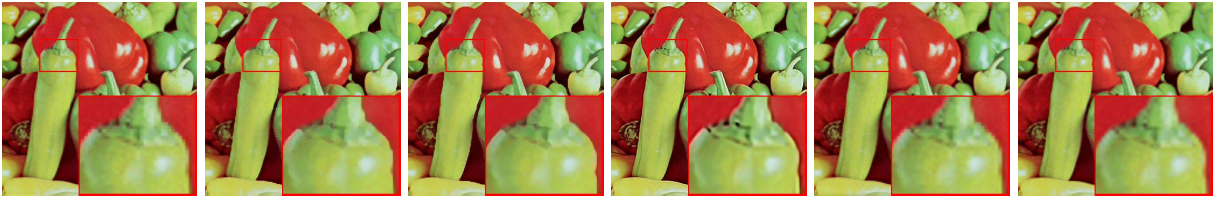}\\
	\caption{SISR results ($\times 3$) of noisy \emph{Peppers}   given by different methods. From left to right: MCI, MCIFD, FD, CRNS, GSE, Lanczos.}\label{reconsimagesnoisenew}
\end{figure*}

\begin{table}[!t]
	\caption{Average  PSNR, SSIM, FSIM, and CC results on noisy test images by different methods ($\times 3$).  (Bold: the best; underline: the second best).}\label{averageQMn}  \vspace{2pt}	\centering \small
	\begin{tabular}{l l l l l l l}
		\hline
		Metric/Method  & MCI &MCIFD&FD &CRNS & GSE & Lanczos  \\
		\hline
		PSNR  & \textbf{28.61} & 27.87  &26.32 & 28.29 & \underline{28.42}&26.50 \\
		SSIM &  \underline{0.7862}& 0.7665 & 0.7392 & \textbf{0.8122} &0.7811&0.7304  \\
		FSIM  & \underline{0.8625} &0.8348& 0.8086& \textbf{0.8802}& 0.8557 &  0.8296\\
		CC & \textbf{0.9733}&0.9682&0.9581&0.9715&\underline{0.9723}&0.9593\\
		\hline
	\end{tabular}
\end{table}

To validate the robustness of MCI algorithm to noise, we conduct experiments on noisy images.
The eight test images and the downsampling procedure are same as the noiseless case. Besides, we add Gaussian noise with a standard deviation $3$ to the LR images. The corresponding quantitative   measurements  for    SISR results are presented in Table \ref{averageQMn} and Figure \ref{reconsimagesnoisenew}.
Analogous to the FD, CRNS, GSE and Lanczos methods,  the quantitative measurements  of performance for MCI decrease slightly  if noise is introduced. This indicates that MCI is not sensitive to noise.

It is noted that  MCI outperforms  the others in PSNR and it is inferior to CRNS in SSIM and FSIM. This is because MCI is effective interpolation method with high accuracy. From Section \ref{S51}, we see that MCI achieves a low mean square error (MSE) in signal reconstruction. Note that low MSE means high PSNR, thus MCI can obtain  high PSNR in image interpolation as well. However, MCI is a 1D interpolation method and therefore it is lack of consideration of 2D  structure of image. Hence MCI does not perform as well as CRNS in SSIM and FSIM.
We summarize  some limitations of MCI in SISR and propose   possible directions for overcoming.
\begin{enumerate}
	\item   MCI deals with the row and column of image separately, which means that it does not consider the 2D structure of image. Therefore, there is a need to develop 2D MCI for reconstructing 2D signals and interpolating 2D images.
	\item MCI does not utilize the information of the blur process. It is known that the blur process can be viewed as a 2D filtering. The main idea of MCI is about reconstructing the original signals  from the samples of filtered signals. Thus it is possible to take blur model into consideration in MCI to improve SISR performance.
	\item MCI is applied to the image interpolation globally. We have mentioned that it is  flexible to control the frequency distribution of reconstructed signal by MCI. Therefore, it is possible to process MCI  locally   to accommodate   different frequency characteristics in  different regions.
	\item There are learning-based and reconstruction-based SISR methods which also  involve interpolation techniques (typically,  bicubic). It is believed that these SISR methods can achieve a improvement if the classical  interpolation method (such as bicubic) is replaced by MCI.
\end{enumerate}

\section{Conclusions}\label{S6}

In this paper, we presented a novel multichannel interpolation for   finite duration signals. We show that the reconstruction of a  continuous  signal using data other than the samples of original signal is feasible.  Under suitable conditions,   only $\mu(I^{\mathbf{N}})$ total number of samples, no matter what their types,  are needed to perfectly recover the  signal of the form (\ref{bandsignal}). Quantitative error analysis for reconstructing non-bandlimited signals is also studied.  Both of the theoretical analysis and experiments show that the proposed interpolation method can effectively restore the original signal along with its Hilbert transform.  Based on FFT, MCI can be implemented   fast and efficiently. A particular application is applied for SISR and the  effectiveness of the proposed algorithm is demonstrated by the experimental studies.    In real applications, nonuniform sampling naturally arises. In view of the advantages of the proposed method, a direction of our future work is to   develop multichannel interpolation for non-uniformly distributed data.



\end{document}